\documentclass[a4paper,fleqn]{cas-dc}
\usepackage{enumitem}
\usepackage[ruled,vlined,linesnumbered]{algorithm2e}
\usepackage{float}
\usepackage{caption}
\usepackage{placeins}
\usepackage{cuted}
\usepackage{graphicx}

\usepackage[T1]{fontenc}
\usepackage[utf8]{inputenc}

\usepackage[numbers]{natbib}

\def\tsc#1{\csdef{#1}{\textsc{\lowercase{#1}}\xspace}}
\tsc{WGM}
\tsc{QE}

\begin{document}
\raggedbottom
\let\WriteBookmarks\relax
\def\floatpagepagefraction{1}
\def\textpagefraction{.001}

\shorttitle{SA-DRL for Ransomware Detection}    

\title[mode=title]{SA-DRL: Security-Aware Deep Reinforcement Learning for Ransomware Detection with Asymmetric Reward Design}

\shortauthors{Ferdous et al.}

\author[1]{Jannatul Ferdous}[orcid=0000-0002-9612-0482]

\ead{jferdous@csu.edu.au}

\credit{Conceptualization, Methodology, Software, Investigation, Writing - original draft}

\author[2]{Rafiqul Islam}

\credit{Supervision, Validation, Writing - review \& editing}

\author[3,4]{Md Zahidul Islam}

\credit{Supervision, Writing - review \& editing}

\affiliation[1]{
    organization={School of Computing, Mathematics and Engineering, Charles Sturt University},
    city={Wagga Wagga},
    state={NSW},
    postcode={2650},
    country={Australia}
}

\affiliation[2]{
    organization={School of Computing, Mathematics and Engineering, Charles Sturt University},
    city={Albury},
    state={NSW},
    postcode={2640},
    country={Australia}
}

\affiliation[3]{
    organization={School of Computing, Mathematics and Engineering, Charles Sturt University},
    city={Bathurst},
    state={NSW},
    postcode={2795},
    country={Australia}
}

\affiliation[4]{
    organization={AI and Cyber Futures Center, Charles Sturt University},
    addressline={Panorama Avenue},
    city={Bathurst},
    state={NSW},
    postcode={2795},
    country={Australia}
}

\cortext[1]{Corresponding author}



\begin{abstract}
Ransomware encrypts victim data in seconds; however, current machine learning detectors use symmetric loss functions that equally penalize missed detections, known as false negatives (FN), and benign false alarms, known as false positives (FP). This assumption is misaligned with operational reality, where an FN causes irreversible data loss and high recovery costs, while an FP is reversible. Detection is further complicated by ransomware variants with diverse behaviors, reducing the effectiveness of fixed supervised weighting strategies. To address these challenges, this study proposes a Security-Aware Deep Reinforcement Learning (SA-DRL) framework that embeds FN--FP cost asymmetry into the reinforcement learning reward signal, optimizing detection policies to minimize missed detections. The framework also introduces a Security-Optimal Model Selection (SOMS) criterion and an adaptive sample-weighting mechanism through episode-level random permutation. Four DRL agents, DQN, DDQN, PPO, and A2C, were trained using a symmetric baseline reward ($R_1$) and a security-aware asymmetric reward ($R_2$) that penalizes FNs more heavily. Each configuration was evaluated with four discount factors, five-fold cross-validation, and three random seeds, resulting in 480 training runs on a balanced dataset. The SOMS criterion prioritizes minimizing the false-negative rate (FNR), maximizing the F1-score, and minimizing training time. Results show that asymmetric reward shaping improves detection performance. The SOMS-selected configuration, DDQN with $R_2$ and $\gamma=0.1$, achieved an FNR of 0.0080, an F1-score of 0.9915, and an AUC of 0.998, reducing missed detections by 67.6\% compared with the best baseline model, MLP with FNR = 0.0247. $R_2$ reduced the mean FNR by 43\% relative to $R_1$ across all configurations. These findings highlight the importance of reward-function design in security-sensitive detection systems. SA-DRL is the first framework to combine runtime features, asymmetric reward design, and a security-first model-selection criterion with statistical validation for ransomware detection.
\end{abstract}



\begin{keywords}
Ransomware detection \sep
Deep reinforcement learning \sep
Asymmetric reward design \sep
False-negative minimization \sep
Behavioral ransomware analysis \sep
DDQN
\end{keywords}

\maketitle

\section{Introduction}

Ransomware has emerged as one of the most operationally destructive and economically consequential categories of malware in the modern threat landscape. Unlike conventional malware that silently exfiltrates data, ransomware weaponizes cryptographic extortion by encrypting victim files and demanding payment, typically in cryptocurrency, to restore access. The impact extends well beyond ransom payment alone. In 2025, the global average cost of ransomware recovery reached \$1.53 million, including downtime, investigation, regulatory exposure, and reputational damage~\cite{ref1}. Ransomware was involved in 44\% of all data breaches analyzed in the Verizon 2025 Data Breach Investigations Report, representing a 12-percentage-point increase over the previous year~\cite{ref2}. Furthermore, global ransomware damage costs are projected to exceed \$57 billion annually in 2025 and surpass \$275 billion by 2031, with attacks occurring approximately every two seconds~\cite{ref3}. Modern ransomware campaigns are increasingly operated through ransomware-as-a-service (RaaS) ecosystems, enabling even low-skilled actors to launch attacks against healthcare systems, financial institutions, and critical infrastructure~\cite{ref4}. In February 2024, the BlackCat attack on Change Healthcare disrupted national prescription processing services and resulted in a reported \$22 million ransom payment, demonstrating the systemic risk ransomware poses to essential services~\cite{ref2}.

Given the operational severity of modern ransomware attacks, substantial research efforts have focused on automated detection mechanisms capable of identifying malicious behavior before encryption is completed. Consequently, ransomware detection has evolved from traditional signature-based approaches and static binary analysis toward supervised machine learning and deep learning models trained on behavioral telemetry~\cite{ref5,ref6}. Recent deep learning architectures have reported detection accuracies exceeding 98\% on benchmark datasets~\cite{ref7,ref8}. However, optimizing primarily for overall classification accuracy conceals a fundamental limitation shared across most prior work: false negatives (FN) and false positives (FP) are typically treated as equally costly during training. This assumption is operationally unrealistic in security-critical environments. A false negative allows ransomware to complete encryption, often resulting in irreversible data loss without the attacker's private key, whereas a false positive usually triggers a reversible quarantine or analyst review. In healthcare environments alone, a single undetected ransomware incident may incur downtime costs approaching \$1.9 million per day~\cite{ref2}. Despite this clear asymmetry, most supervised ransomware detectors rely on symmetric loss functions, such as binary cross-entropy, which cannot directly encode security-sensitive cost imbalance into the optimization objective~\cite{ref9,ref10}.

Beyond cost asymmetry, ransomware detection is further complicated by the increasingly evasive and behaviorally adaptive nature of contemporary ransomware families. Dynamic behavioral analysis approaches that monitor API calls, registry activity, file-system operations, entropy changes, and network behavior provide substantially greater resilience against obfuscation than static methods because malicious runtime behavior remains relatively consistent even when binary structures change~\cite{ref11,ref12}. Deep learning models trained on behavioral representations have demonstrated strong performance, with convolutional, recurrent, and transformer-based architectures reporting accuracies in the 96--99\% range~\cite{ref13,ref14,ref15}. Nevertheless, these approaches still optimize aggregate classification performance under symmetric training objectives. Zahoora et al.~\cite{ref9} partially addressed this issue using post-hoc cost-sensitive ensemble weighting, while Deng et al.~\cite{ref16} demonstrated the feasibility of deep reinforcement learning (DRL) for ransomware classification using static PE header features. However, to the best of our knowledge, no prior work has embedded FN--FP cost asymmetry directly into the reward signal of a DRL agent operating on behavioral runtime features.

These limitations indicate that improving ransomware detection requires not only stronger behavioral representations, but also learning mechanisms capable of incorporating operational security priorities directly into the optimization process. Deep Reinforcement Learning (DRL) provides a principled framework for addressing this challenge. In DRL, an agent learns a policy through interaction with an environment and cumulative reward feedback, enabling the reward function itself to encode domain-specific operational priorities. Prior studies in adjacent cybersecurity domains have demonstrated the effectiveness of reward shaping for security-sensitive decision making. Nguyen and Reddi~\cite{ref17} established theoretical foundations for DRL in cybersecurity, showing that reward engineering substantially influences learned agent behavior. Mohamed and Ejbali~\cite{ref18} applied deep SARSA to network anomaly detection, while REACT-D3QN~\cite{ref19} employed asymmetric FN/FP penalties to enforce minimum detection guarantees in intrusion detection systems. Ibrahim et al.~\cite{ref20} further demonstrated that asymmetric reward engineering consistently improves performance in safety-critical classification environments. Despite these advances, the application of DRL to behavioral ransomware detection, particularly the interaction between asymmetric reward design, discount factor selection, and agent architecture, remains substantially underexplored.

To address these gaps, this study proposes a Security-Aware Deep Reinforcement Learning (SA-DRL) framework for binary ransomware detection. A custom Gymnasium-compatible Markov Decision Process (MDP) environment was constructed using a balanced behavioral dataset containing 2,000 samples, comprising 1,000 ransomware instances across 30 families and 1,000 benign samples represented through runtime behavioral features. Two reward functions were designed and evaluated: a baseline symmetric reward (R1: $\pm1.0$) and a security-sensitive asymmetric reward (R2) that applies a penalty of $-2.0$ for false negatives and $-0.5$ for false positives, reflecting a ratio of operational costs FN-to-FP 4:1. Four DRL agents --- DQN, DDQN, PPO, and A2C were evaluated across four discount factors ($\gamma \in \{0.1, 0.5, 0.9, 0.99\}$), five-fold stratified cross-validation, and three random seeds under identical experimental conditions, resulting in 480 controlled training runs. In addition, this study introduces a Security-Optimal Model Selection (SOMS) criterion that prioritizes minimum FNR, maximum F1-score, and minimum training time as a formal security-first selection protocol. The SOMS-selected configuration, DDQN with R2 and $\gamma = 0.1$, achieved an FNR of 0.0080 and an F1-score of 0.9915, reducing missed ransomware detections by 67.6\% relative to the best supervised baseline (MLP, FNR = 0.0247). All performance differences were statistically validated using Friedman and Wilcoxon signed-rank tests.

The key contributions of this study are as follows:
\begin{enumerate}
    \item \textbf{Asymmetric Reward Design for Security-Aware Ransomware Detection:} We propose a DRL framework in which asymmetric misclassification costs are embedded directly into the reward signal, enabling the agent to learn a security-aware decision boundary through reward-shaped optimization rather than post-hoc threshold adjustment.

    \item \textbf{Comprehensive Multi-Factor Comparative Evaluation:} We present a controlled comparison of four DRL algorithms such as DQN, DDQN, PPO, and A2C across two reward functions, four discount factors, five folds, and three random seeds, resulting in 480 experimentally controlled training runs.

    \item \textbf{Security-Optimal Model Selection (SOMS):} We introduce a formal security-first model-selection criterion prioritizing minimum FNR, maximum F1-score, and minimum training time for high-risk binary classification environments.

    \item \textbf{Adaptive Sample Weighting Through Episode-Level Permutation:} The proposed framework introduces an implicit adaptive weighting effect through episode-level random permutation, improving exposure to behaviorally complex samples without explicit per-sample weighting computation.

    \item \textbf{Statistically Validated Performance Superiority:} Experimental results validated through Friedman and Wilcoxon signed-rank tests demonstrate that the SOMS-selected DDQN configuration significantly outperformed all evaluated baselines while substantially reducing missed ransomware detections.
\end{enumerate}

The remainder of this paper is organized as follows. Section~2 reviews related work on ransomware detection and DRL-based cybersecurity approaches. Section~3 presents the proposed methodology. Section~4 describes the experimental setup. Section~5 reports and analyzes the experimental results. Section~6 discusses the findings, limitations, and implications of the study. Finally, Section~7 concludes the paper and outlines future research directions.

\section{Related Work}
\label{sec:related_work}

This section reviews four streams of literature that frame the present study: signature-based and static ransomware detection, machine learning and deep learning using behavioral features, deep reinforcement learning for cybersecurity, and the research gap addressed by this work.

\subsection{Signature-Based and Static Analysis}
\label{subsec:static_analysis}

Early ransomware detection systems relied primarily on signature matching, where file hashes, byte sequences, and known malicious strings were compared against threat databases. Although computationally efficient and straightforward to deploy, these systems are inherently reactive because they can only detect previously catalogued variants. They are also easily bypassed when structural modifications alter the binary signature without changing the underlying malicious logic~\cite{ref21}. This limitation is critical because modern ransomware families frequently employ polymorphic recompilation and automated obfuscation to generate distinct binaries across victims~\cite{ref22}.

Static analysis attempts to address this limitation by extracting features from executable structures without executing the sample. Common techniques include opcode n-gram extraction~\cite{ref23}, entropy-based packing detection~\cite{ref24}, portable executable (PE) header analysis~\cite{ref25}, and binary visualization using convolutional classifiers~\cite{ref26}. These methods offer speed advantages and often achieve high accuracy on closed-world benchmarks. Recent PE-header-based frameworks using DLL lists, section entropy, and embedded strings have reported strong ransomware detection performance~\cite{ref6}, while PE-header-based machine learning with SHAP-based feature attribution has achieved accuracy above 97\%~\cite{ref5}.

However, the limitations of static analysis are well established. Moser et al.~\cite{ref21} showed that strong obfuscation can randomize static signatures and features without altering program semantics. Contemporary ransomware exploits this weakness through code obfuscation, polymorphic recompilation, runtime packing, process hollowing, and sandbox-detection techniques~\cite{ref4,ref27,ref28}. Deng et al.~\cite{ref16} further showed that even a DRL-based ransomware detector using PE header features remains exposed to binary-level mutation. Recent reviews of AI-based ransomware detection also confirm that static approaches generally underperform dynamic and hybrid methods against zero-day and polymorphic variants~\cite{ref29,ref30,ref31}. These limitations have motivated a broader transition toward runtime behavioral analysis, where malicious intent is inferred from observable system interactions rather than executable structure alone.

\subsection{Machine Learning and Deep Learning with Behavioral Features}
\label{subsec:behavioral_ml_dl}

The vulnerability of static analysis to adversarial obfuscation has encouraged extensive research on machine learning models trained on dynamic behavioral features collected during sandboxed or monitored execution. Behavioral features are more resistant to evasion than static byte patterns because ransomware must eventually perform observable malicious actions, such as enumerating, accessing, encrypting, and deleting files, regardless of how the binary is packed. Common behavioral features include API call frequencies and sequences, registry modification counts, file-system operations, entropy changes, network connection patterns, and mutex activity~\cite{ref31}.

Classical machine learning approaches using behavioral features have achieved strong baseline performance. Berrueta et al.~\cite{ref31} developed a 50-feature dataset based on file access, network behavior, and system calls, reporting near-perfect accuracy using Random Forest and Gradient Boosted Trees. Khan et al.~\cite{ref32} proposed a Digital DNA Sequencing engine based on API calls and Random Forest, achieving 97.3\% accuracy. Other studies have also demonstrated the effectiveness of behavioral machine learning for ransomware classification~\cite{ref13,ref14}. Although these methods are efficient and interpretable, they treat detection as a static classification problem and therefore lack adaptability to evolving threat behavior and changing operational cost requirements.

Deep learning methods extend this capability by modelling temporal and sequential dependencies in behavioral traces. Architectures such as LSTM, RNN, CNN, BERT, and hybrid models have achieved detection accuracies in the 96--99\% range using ransomware behavioral telemetry~\cite{ref10,ref15,ref33}. These models are particularly suitable for API call sequences because such sequences contain ordering patterns that cannot be fully captured by flat feature vectors.

Despite these advances, most ML and DL detectors treat all training samples and error types uniformly. Consequently, they may struggle under behavioral heterogeneity, where modern ransomware exhibits partially benign or evasive runtime patterns. More importantly, these models typically minimize symmetric loss functions that assign equal penalties to false negatives and false positives. The cost-sensitive Pareto Ensemble Classifier proposed by Zahoora et al.~\cite{ref32} partially addressed this issue through post-hoc cost weighting of pre-trained classifiers. However, this strategy does not embed asymmetric security costs directly into the learning objective. This limitation motivates the DRL-based formulation proposed in this study, where FN--FP cost asymmetry is encoded directly into the reward signal from the beginning of training.

\subsection{Deep Reinforcement Learning for Cybersecurity}
\label{subsec:drl_cybersecurity}

Deep Reinforcement Learning (DRL) has gained increasing attention in cybersecurity because many security tasks involve asymmetric decision costs and continuously evolving threat patterns. Unlike supervised classifiers trained on fixed objective functions, DRL agents learn through interaction with an environment and update their policies using reward feedback. This makes DRL suitable for domains where the optimization objective must reflect operational priorities rather than aggregate accuracy alone~\cite{ref22,ref34}. By formulating detection as a Markov Decision Process (MDP), the reward function can encode domain-specific cost asymmetries and guide the agent toward security-oriented decision policies.

DRL has been widely explored in network intrusion detection, where value-based agents often outperform policy-gradient methods on discrete classification benchmarks. Prior studies have also shown that lower discount factors can improve detection performance in security-oriented classification tasks~\cite{ref35}, while A2C may exhibit higher false-negative rates under adversarial perturbation~\cite{ref36}. In malware classification, DDQN has shown consistent advantages over DQN and supervised baselines because it reduces Q-value overestimation bias~\cite{ref37}. Deng et al.~\cite{ref16} demonstrated the feasibility of DRL for binary ransomware classification using PE header features, confirming that MDP-based formulations can be applied to ransomware detection.

Reward design is particularly important in security-sensitive DRL systems. The REACT-D3QN framework~\cite{ref19} used asymmetric FN/FP penalties to enforce a minimum guaranteed detection rate rather than optimizing aggregate accuracy. Bates et al.~\cite{ref38} further showed that symmetric reward structures can produce suboptimal detection policies in threat environments. However, despite these advances, no prior DRL study has embedded asymmetric FN--FP cost asymmetry into the reward signal of a behavioral ransomware detector.

\subsection{Research Gap Summary}
\label{subsec:research_gap}

The reviewed literature demonstrates substantial progress in behavioral ransomware detection and DRL-based cybersecurity learning, while also revealing several unresolved methodological gaps. First, most ransomware detectors rely on symmetric loss functions that do not reflect the operational cost imbalance between false negatives and false positives. Second, existing DRL-based ransomware detection has not embedded asymmetric FN--FP costs directly into the reward signal using behavioral runtime features. Third, no systematic multi-agent DRL comparison has evaluated ransomware detection under paired symmetric and asymmetric reward conditions. Fourth, the discount factor $\gamma$ has not been examined as a security-relevant hyperparameter for ransomware detection. Finally, existing studies typically select models using aggregate accuracy or F1-score, rather than a formal security-first criterion. The proposed SA-DRL framework addresses these gaps by combining behavioral runtime features, asymmetric reward design, multi-agent DRL evaluation, discount-factor analysis, and Security-Optimal Model Selection (SOMS).

\section{Methodology}
\label{sec:methodology}

\begin{figure*}[h]
\centering
\includegraphics[width=\textwidth,height=.80\textheight,keepaspectratio]{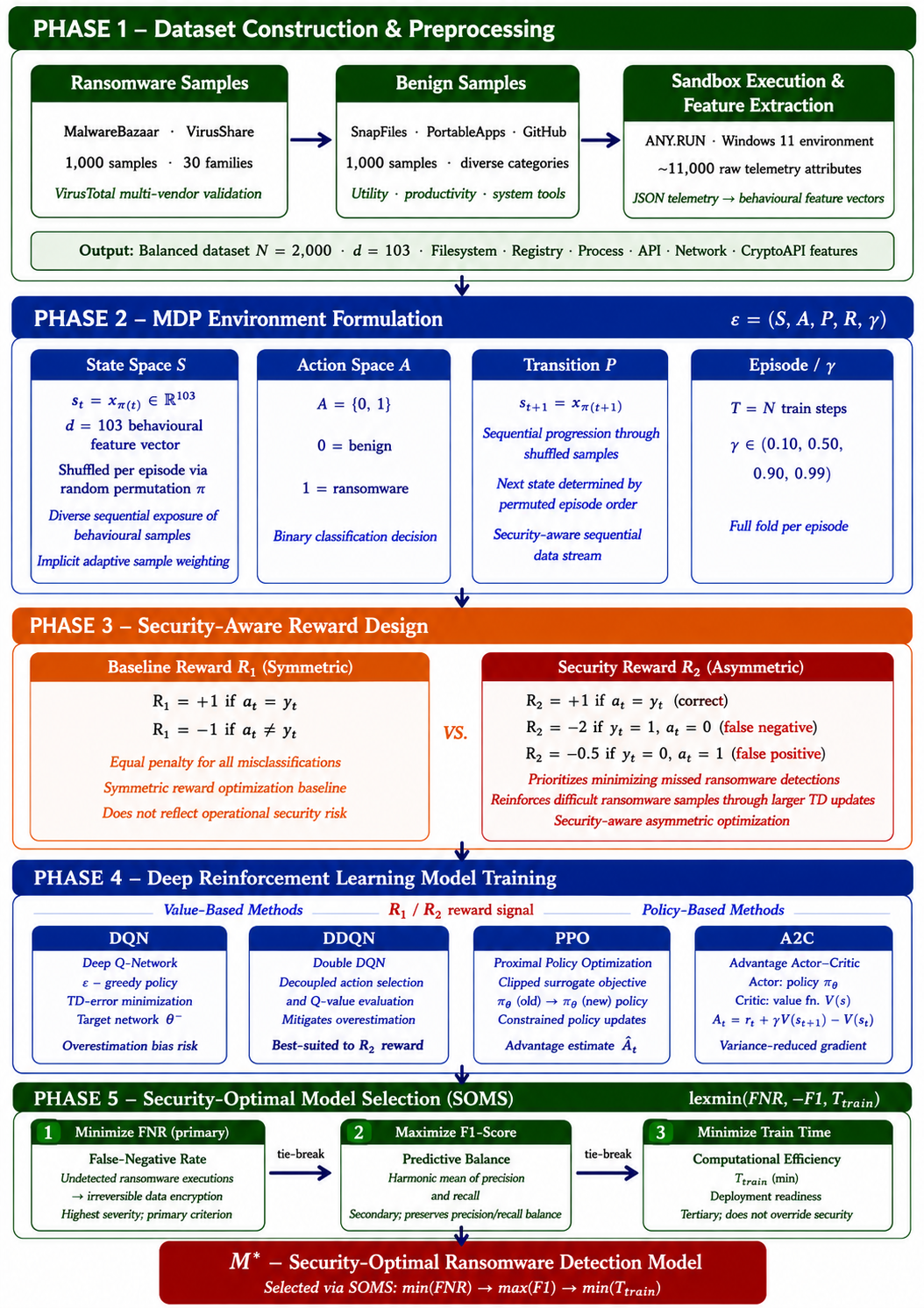}
\caption{Workflow of the proposed SA-DRL framework for ransomware detection. The pipeline begins with balanced behavioral features, applies episode-level random permutation, constructs state representations, and trains DRL agents under symmetric and asymmetric reward settings. Following sequential state transitions and episode termination, model performance is evaluated, and the final configuration is selected using the proposed Security-Optimal Model Selection (SOMS) criterion.}
\label{fig:framework}
\end{figure*}

To address the limitations identified in Section~\ref{sec:related_work}, this section presents the complete methodological pipeline of the proposed Security-Aware Deep Reinforcement Learning (SA-DRL) framework for ransomware detection. The framework consists of five tightly integrated components: dataset construction and preprocessing (Section~\ref{subsec:dataset}); formal Markov Decision Process (MDP) formulation and environment design (Section~\ref{subsec:mdp}), including an adaptive sample-weighting mechanism induced through episode-level random permutation; security-aware asymmetric reward design (Section~\ref{subsec:reward_design}); deep reinforcement learning model architectures (Section~\ref{subsec:drl_models}); and the proposed Security-Optimal Model Selection (SOMS) criterion (Section~\ref{subsec:soms}). In addition, baseline model specifications and the statistical validation protocol are presented in Sections~\ref{subsec:baseline_models} and~\ref{subsec:statistical_validation}, respectively. All mathematical formulations were derived directly from the implemented framework and were fully reproducible. The overall workflow of the proposed SA-DRL framework is illustrated in Fig.~\ref{fig:framework}.

\subsection{Dataset Construction and Preprocessing}
\label{subsec:dataset}

To evaluate security-aware ransomware detection under realistic conditions, this study uses a custom balanced behavioral dataset comprising 2,000 Windows executable samples: 1,000 ransomware samples across 30 families and 1,000 benign samples. The dataset was constructed to reflect contemporary Windows 11 environments and to address limitations commonly observed in legacy Cuckoo-based corpora.

\subsubsection{Sample Collection}
\label{subsubsec:sample_collection}

Ransomware samples were collected from MalwareBazaar~\cite{ref39} and VirusShare~\cite{ref40}, covering 30 distinct ransomware families. The ransomware families were selected using two criteria: (i) threat intelligence reports highlighted their high frequency and substantial global organizational impact during 2019--2024~\cite{ref41,ref42,ref43,ref44,ref45,ref46}; and (ii) the family classification was confirmed by at least two independent reports from reputable cybersecurity firms. Based on this strategy, prominent families such as LockBit, Conti, BlackCat, MedusaLocker, Phobos, and WannaCry were included to capture diverse encryption strategies, persistence mechanisms, and command-and-control behaviors.

Individual ransomware samples were selected following the methodology in~\cite{ref25}, using three criteria based on VirusTotal vendor-engine detections. A sample was retained as ransomware only if it satisfied all three conditions: (i) detection as malicious by at least 45 antivirus engines; (ii) explicit identification as ransomware by at least 15 engines; and (iii) majority agreement on the ransomware family classification. This multi-vendor validation strengthened labeling confidence and dataset integrity. Table~\ref{tab:ransomware_families} reports the ransomware family distribution.

Benign samples were collected from trusted software repositories, including SnapFiles~\cite{ref47}, PortableApps.com~\cite{ref48}, and GitHub~\cite{ref49}, ensuring diversity across utility tools, productivity software, and system applications.

\begin{table}[!h]
\caption{Ransomware families included in this study with their sample counts. Families were selected based on global prevalence, documented impact, and multi-vendor confirmation of their classification.}
\label{tab:ransomware_families}
\centering
\footnotesize
\setlength{\tabcolsep}{4pt}

\begin{tabular}{lr|lr}
\toprule
\textbf{Family name} & \textbf{No.} &
\textbf{Family name} & \textbf{No.} \\
\midrule

LockBit & 114 & Phobos & 22 \\
GandCrab & 97 & WannaCry & 21 \\
MedusaLocker & 63 & WastedLocker & 20 \\
NetWalker & 56 & BlackMatter & 20 \\
Conti & 55 & BlackBasta & 20 \\
Babuk & 52 & Ryuk & 20 \\
Cerber & 49 & RagnarLocker & 19 \\
Maze & 47 & Mespinoza & 19 \\
REvil & 43 & AvosLocker & 15 \\
DarkSide & 41 & Avadon & 14 \\
Dharman & 36 & MountLocker & 11 \\
Thanos & 31 & Exorcist & 11 \\
TeslaCrypt & 31 & Mallox & 10 \\
Makop & 23 & Nefilim & 10 \\
Akira & 22 & BlueSky & 8 \\

\midrule

\multicolumn{2}{l|}{\textbf{Total families}} &
\multicolumn{2}{r}{\textbf{30}} \\

\multicolumn{2}{l|}{\textbf{Total samples}} &
\multicolumn{2}{r}{\textbf{1000}} \\

\bottomrule
\end{tabular}
\end{table}

\begingroup
\setlength{\abovedisplayskip}{3pt}
\setlength{\belowdisplayskip}{3pt}
\setlength{\abovedisplayshortskip}{1.5pt}
\setlength{\belowdisplayshortskip}{1.5pt}

\subsubsection{Sandbox Execution and Feature Engineering}
\label{subsubsec:sandbox_execution}

All samples were executed in the ANY.RUN interactive sandbox~\cite{ref50} under a Windows~11 environment with native network connectivity. ANY.RUN was selected because it supports modern operating systems, provides rich behavioral telemetry, and demonstrates improved resistance to sandbox-evasion techniques compared with legacy Cuckoo-based systems~\cite{ref51}. Each execution generated a JSON report containing approximately 11,000 raw telemetry attributes.

The raw JSON telemetry generated by ANY.RUN was processed using a custom Python-based extraction pipeline to construct a structured behavioral feature matrix. For each sample, the extraction framework parsed the JSON report and derived behavioral indicators across multiple categories, including filesystem operations (file creation, deletion, renaming, and encryption-event counts), registry modifications (key reads, writes, and deletions), process activity, API-call frequencies, network behavior (DNS-query counts, HTTP/HTTPS connection attempts, and command-and-control indicators), and CryptoAPI usage.

This process produced an initial behavioral dataset represented as:
\begin{equation}
X \in \mathbb{R}^{N \times d}
\label{eq:dataset_matrix}
\end{equation}

where $N = 2000$ denotes the total number of executable samples and $d$ represents the number of extracted behavioral features.

The corresponding binary label vector is defined as:
\begin{equation}
y =
[\underbrace{1,\ldots,1}_{1000},
\underbrace{0,\ldots,0}_{1000}]^{T}
\in \{0,1\}^{2000}
\label{eq:label_vector}
\end{equation}

where ransomware samples are assigned label 1 and benign samples are assigned label 0. The balanced 1:1 class distribution eliminates majority-class exploitation bias from all reported evaluation metrics.

\begingroup
\setlength{\abovedisplayskip}{3pt}
\setlength{\belowdisplayskip}{3pt}
\setlength{\abovedisplayshortskip}{1.5pt}
\setlength{\belowdisplayshortskip}{1.5pt}

\subsubsection{Preprocessing}
\label{subsubsec:preprocessing}

To ensure reproducibility and reduce feature redundancy, domain-guided feature engineering was applied, including bias removal, categorical encoding, and feature standardization. This process reduced the initial telemetry representation to a final behavioral feature set suitable for both machine learning and reinforcement learning models.

All features were standardized using zero-mean, unit-variance normalization. To prevent data leakage during cross-validation, the \texttt{StandardScaler} was fitted exclusively on the training fold and subsequently applied to both training and validation partitions as follows:
\begin{equation}
\widetilde{x}_{ij}
=
\frac{x_{ij} - \mu_{j}^{(k)}}
{\sigma_{j}^{(k)}}
\label{eq:standardization}
\end{equation}
where $\mu_{j}^{(k)}$ and $\sigma_{j}^{(k)}$ denote the mean and standard deviation of feature $j$ computed exclusively from training fold $k$. All normalized values were cast to \texttt{float32} precision to match the numerical input requirements of the PyTorch policy networks.

The final output consisted of a 103-dimensional behavioral feature vector suitable for ML- and RL-based ransomware detection. The resulting dataset provides balanced class distribution, family diversity, modern operating-system fidelity, and strong labeling reliability, making it suitable for evaluating ransomware detection performance under realistic threat conditions.

After constructing the behavioral feature representation, the ransomware detection problem was formally modeled as a sequential decision-making process within a reinforcement learning environment.

\subsection{Problem Formulation and Reinforcement Learning Environment Design}
\label{subsec:mdp}

The behavior-based ransomware detection task is formulated as a finite-horizon Markov Decision Process (MDP), where each episode processes all training samples exactly once in a shuffled sequence and terminates after a fixed number of steps. The sequential decision-making process is formally defined as:
\begin{equation}
\mathcal{M} = (\mathcal{S}, \mathcal{A}, P, R, \gamma, T)
\label{eq:mdp_tuple}
\end{equation}
where $\mathcal{S}$ denotes the state space, $\mathcal{A}$ is the action space, $P$ is the transition function, $R$ is the reward function, $\gamma$ is the discount factor, and $T$ is the finite episode horizon. The environment was constructed from a balanced dataset of $N=2000$ behavioral samples, comprising 1,000 ransomware and 1,000 benign samples. Each sample is represented by a $d=103$ dimensional behavioral feature vector, resulting in a binary classification problem over runtime behavioral telemetry.

\subsubsection{State Space}
\label{subsubsec:state_space}

The ransomware dataset is defined as:
\begin{equation}
\mathcal{D} = \{(x_i, y_i)\}_{i=1}^{N}
\label{eq:dataset}
\end{equation}
where $x_i \in \mathbb{R}^{d}$ denotes the feature vector of the $i$-th sample, and $y_i \in \{0,1\}$ denotes its class label. Labels were assigned according to the dataset ordering as follows:
\begin{equation}
y_i =
\begin{cases}
1, & 1 \leq i \leq 1000 \quad \text{(ransomware)},\\
0, & 1001 \leq i \leq 2000 \quad \text{(benign)}.
\end{cases}
\label{eq:label_assignment}
\end{equation}
At each time step $t$, the agent observes the current behavioral feature vector:
\begin{equation}
s_t = x_{\pi(t)} \in \mathbb{R}^{103}
\label{eq:state_definition}
\end{equation}
where $\pi$ denotes a random permutation of the training indices applied at the beginning of each episode. Accordingly, the state space is defined as:
\begin{equation}
\mathcal{S} \subseteq \mathbb{R}^{103}
\label{eq:state_space}
\end{equation}

\subsubsection{Action Space}
\label{subsubsec:action_space}

The action space is binary:
\begin{equation}
\mathcal{A} = \{0,1\}
\label{eq:action_space}
\end{equation}
where action 0 denotes benign and action 1 denotes ransomware. Therefore, each action directly corresponds to a classification decision made by the agent.

\subsubsection{Random Permutation and Episode Initialization}
\label{subsubsec:episode_initialization}

A key component of the environment design is episode-level randomization of the sample order. At the beginning of each episode, the training samples are randomly reordered before being presented to the agent. This operation is represented by a permutation:
\begin{equation}
\pi: \{1,\ldots,N_{\mathrm{train}}\} \rightarrow \{1,\ldots,N_{\mathrm{train}}\}
\label{eq:permutation}
\end{equation}
where $\pi$ is a random permutation of the sample indices. A permutation is a reordering of the index set such that every sample appears exactly once and no sample is repeated or omitted. After applying $\pi$, the training sequence becomes:
\begin{equation}
\bigl(x_{\pi(1)}, y_{\pi(1)}\bigr),
\bigl(x_{\pi(2)}, y_{\pi(2)}\bigr),
\ldots,
\bigl(x_{\pi(N_{\mathrm{train}})}, y_{\pi(N_{\mathrm{train}})}\bigr)
\label{eq:permuted_sequence}
\end{equation}
Thus, every sample appears once per episode, but the order changes at every episode. Random permutation is important for four reasons: it removes dependence on fixed dataset ordering, reduces order-induced learning bias, ensures complete exposure to the training set while preserving sample-label correspondence, and provides an implicit adaptive sample-weighting effect. By varying sample sequences, the asymmetric reward signal reinforces correct classification of difficult ransomware samples without requiring explicit per-sample weight computation. This is particularly important under the asymmetric reward function, where the larger false-negative penalty produces stronger temporal-difference updates for missed ransomware samples and aligns policy learning with the operational security objective.

\subsubsection{Transition Dynamics}
\label{subsubsec:transition_dynamics}

Once an episode starts and the shuffled sequence is fixed, the environment advances deterministically from one sample to the next. After the agent observes:
\begin{equation}
s_t = x_{\pi(t)}
\label{eq:current_state}
\end{equation}
and selects action $a_t$, the next state is defined as:
\begin{equation}
s_{t+1} = x_{\pi(t+1)}
\label{eq:next_state}
\end{equation}
Hence, the transition probability is:
\begin{equation}
P(s_{t+1} \mid s_t, a_t) =
\begin{cases}
1, & s_{t+1} = x_{\pi(t+1)},\\
0, & \text{otherwise}.
\end{cases}
\label{eq:transition_probability}
\end{equation}
The agent's prediction does not alter the next sample presented by the environment. Instead, the environment behaves as a shuffled sequential data stream. Therefore, reinforcement learning is used to optimize decision-making under a reward signal rather than to control future environmental evolution. This formulation is suitable for static behavioral datasets, where the main objective is security-aware decision optimization.

\subsubsection{Episodic Structure}
\label{subsubsec:episodic_structure}

Each episode corresponds to a complete pass over the current training fold. Let $T=N_{\mathrm{train}}$ denote the number of training samples in a given cross-validation split. The episode proceeds as:
\begin{equation}
t = 1,2,\ldots,T
\label{eq:episode_steps}
\end{equation}
At each step, the environment returns the current shuffled sample, receives a binary action, computes the reward, and advances to the next sample. The decision process is therefore finite and terminates when all samples have been processed:
\begin{equation}
\mathrm{terminated} =
\begin{cases}
\mathrm{True}, & t \geq T,\\
\mathrm{False}, & t < T.
\end{cases}
\label{eq:termination}
\end{equation}
After termination, the environment returns a zero-valued dummy observation:
\begin{equation}
s_{T+1} = \mathbf{0} \in \mathbb{R}^{103}
\label{eq:dummy_state}
\end{equation}
This finite-horizon episodic structure ensures that each training episode has a well-defined beginning, endpoint, and complete coverage of the available training data.

\subsubsection{Return Objective}
\label{subsubsec:return_objective}

The agent maximizes the expected discounted cumulative return:
\begin{equation}
J(\pi_{\theta}) =
\mathbb{E}_{\pi_{\theta}}
\left[
\sum_{k=0}^{T-t} \gamma^{k} r_{t+k}
\right]
\label{eq:return_objective}
\end{equation}
where $\pi_{\theta}$ denotes the parameterized policy and $r_{t+k}$ is the reward received at future step $t+k$.

\subsubsection{Discount Factor}
\label{subsubsec:discount_factor}

The discount factor controls the relative importance of immediate and future rewards. Four values are evaluated as security-relevant hyperparameters across all DRL algorithms:
\begin{equation}
\gamma \in \{0.10, 0.50, 0.90, 0.99\}
\label{eq:discount_values}
\end{equation}
Overall, the proposed RL environment models ransomware detection as a sequential classification problem based on behavioral feature states. Each episode begins by randomly shuffling the training samples so that the agent observes every sample exactly once in a different order. The agent then processes the shuffled sequence step by step, selects a binary class label for each state, and receives a reward based on the correctness and security implications of that decision. This formulation preserves the advantages of reinforcement learning for policy optimization while remaining faithful to the static and sample-based nature of the ransomware dataset.

\subsubsection{Adaptive Sample Weighting via DRL-Driven Episode Permutation}
\label{subsubsec:adaptive_sample_weighting}

Conventional supervised learning assigns equal importance to all samples within each epoch, which is suboptimal for ransomware detection because behaviorally ambiguous samples carry higher security risk. The proposed SA-DRL framework addresses this limitation through an implicit adaptive sample-weighting effect that emerges from the interaction between the episodic MDP structure, episode-level random permutation, and asymmetric reward design.

At the beginning of each episode, the training set is randomly permuted so that no sample occupies a fixed temporal position during training. The agent sequentially processes the permuted samples and updates the value estimates using the temporal-difference (TD) error:
\begin{equation}
\delta_t =
r_t + \gamma \max_{a'} Q(s_{t+1}, a'; \theta^{-})
- Q(s_t, a_t; \theta)
\label{eq:td_error}
\end{equation}
The corresponding parameter update is:
\begin{equation}
\theta \leftarrow \theta +
\alpha \delta_t \nabla_{\theta} Q(s_t, a_t; \theta)
\label{eq:q_update}
\end{equation}
where $\alpha$ is the learning rate, $\theta$ denotes the online network parameters, and $\theta^{-}$ denotes the target network parameters. Under the asymmetric reward function, false negatives receive the largest penalty, producing larger TD errors and correspondingly stronger parameter updates than false positives or correct classifications. Consequently, difficult ransomware samples that repeatedly trigger high-penalty updates receive greater training emphasis across diverse episode contexts, functionally resembling adaptive sample weighting without explicit per-sample weight computation.

Unlike explicit reweighting methods, such as class-weighted loss or focal loss, the proposed mechanism requires no manual weight specification or additional computational overhead. Instead, adaptive weighting emerges naturally from the stochastic episode structure and asymmetric reward-driven learning dynamics.

\subsection{Security-Aware Reward Design}
\label{subsec:reward_design}

Reward design is critical in reinforcement learning because it determines how an agent evaluates classification decisions. In conventional classification settings, all misclassifications are typically treated as equally costly. However, this assumption is inappropriate for ransomware detection. False negatives, corresponding to missed ransomware samples, can lead to severe consequences such as data encryption, financial loss, and system compromise. In contrast, false positives primarily incur operational overhead through alerts, quarantine actions, or manual inspection. To address this asymmetry, this study introduces two reward formulations: a baseline symmetric reward (R1) and a security-aware asymmetric reward (R2).

\subsubsection{Baseline Reward Function}
\label{subsubsec:baseline_reward}

The baseline reward function assigns equal importance to correct and incorrect predictions:
\begin{equation}
R_1(a_t,y_t)=
\begin{cases}
+1, & a_t = y_t,\\
-1, & a_t \neq y_t.
\end{cases}
\label{eq:reward_r1}
\end{equation}
where $a_t$ denotes the predicted class and $y_t$ denotes the true class at time step $t$. This formulation reflects standard reinforcement learning classification settings, where all errors are penalized equally regardless of their operational impact.

\subsubsection{Security-Aware Asymmetric Reward Function}
\label{subsubsec:asymmetric_reward}

To better reflect operational security priorities in ransomware detection, the proposed reward function $R_2$ applies asymmetric penalties according to the type of classification outcome:
\begin{equation}
R_2(a_t,y_t)=
\begin{cases}
+1, & a_t = y_t \quad \text{(correct: TP or TN)},\\
-2.0, & y_t = 1 \ \text{and} \ a_t = 0 \quad \text{(false negative)},\\
-0.5, & y_t = 0 \ \text{and} \ a_t = 1 \quad \text{(false positive)}.
\end{cases}
\label{eq:reward_r2}
\end{equation}
This reward structure is directly aligned with the implemented environment: correct classifications are rewarded, false negatives receive the largest penalty, and false positives receive a smaller penalty.

The false-negative penalty ($-2.0$) is four times larger than the false-positive penalty ($-0.5$), encoding the operational asymmetry that missing ransomware is substantially more costly than raising a false alarm. This ratio is deliberately conservative relative to real-world estimates. A single undetected ransomware execution may incur average recovery costs of \$1.53 million, and healthcare downtime may reach \$1.9 million per day~\cite{ref1,ref2}. By contrast, a false positive typically generates a reversible quarantine alert requiring analyst review. Therefore, the 4:1 ratio encodes the correct security-critical direction without claiming a precise empirical cost ratio or introducing excessive reward magnitude that may destabilize Q-value learning. Specifically, switching from R1 to R2 reduces the false-negative rate by 43\%, while increasing the false-positive rate by only 0.20 percentage points. Prior DRL-based security work has used more aggressive ratios; for example, Haddane et al.~\cite{ref18} used a 10:1 asymmetric penalty for cross-dataset intrusion detection under severe concept drift. In contrast, the more conservative 4:1 ratio is appropriate for the controlled, balanced, single-distribution ransomware dataset used in this study. Practitioners deploying this framework under different class distributions or threat profiles should recalibrate this ratio accordingly, as discussed in Section~\ref{sec:discussion}.

The effective cost-pressure differential between false negatives and false positives is:
\begin{equation}
\frac{|\Delta_{\mathrm{FN}}|}{|\Delta_{\mathrm{FP}}|}
=
\frac{|-2.0-(+1.0)|}{|-0.5-(+1.0)|}
=
\frac{3.0}{1.5}
=
2.0
\label{eq:cost_pressure}
\end{equation}
This indicates that the effective optimization pressure toward avoiding false negatives is approximately twice that toward avoiding false positives.

Under $R_1$, the agent is encouraged to maximize overall classification accuracy. In contrast, under $R_2$, the learning objective becomes explicitly security-aware because the reward structure assigns a larger penalty to false negatives than to false positives. Consequently, maximizing $J(\pi_{\theta})$ under $R_2$ encourages policies that minimize missed ransomware detections, even at the cost of a slight increase in false positives. This trade-off is appropriate for ransomware defense because the cost of undetected execution far exceeds the cost of additional analyst review.

Once the security-aware reward structure was defined, multiple DRL algorithms were evaluated to investigate how different learning paradigms respond to asymmetric security-driven optimization objectives.

\subsection{Deep Reinforcement Learning Model Training}
\label{subsec:drl_models}

Deep Reinforcement Learning (DRL) combines reinforcement learning with deep neural networks to enable autonomous decision-making in high-dimensional environments. As illustrated in Fig.~\ref{fig:drl_framework}, the DRL agent in the proposed framework functions as a sequential classifier. At each time step, the agent receives a 103-dimensional behavioral feature vector representing an executable sample, processes it through a neural network, and selects a binary action corresponding to either benign or ransomware classification. The environment then returns a reward signal encoding both the correctness and security implications of the decision, allowing the agent to iteratively refine its policy through cumulative reward optimization over an episode.

Unlike supervised classifiers that optimize a fixed loss function across all training samples simultaneously, DRL agents update their parameters through sequential interaction with the environment. This enables the reward function to encode asymmetric operational costs directly into the learning process.

To evaluate the proposed framework, four DRL algorithms were investigated: DQN, DDQN, PPO, and A2C. These algorithms represent the two major DRL paradigms: value-based learning and policy-gradient learning. All models were trained under the same MDP formulation and reward settings defined in Sections~\ref{subsec:mdp} and~\ref{subsec:reward_design}, ensuring fair and consistent comparison across all experimental conditions.
\begin{center}
\includegraphics[
width=\columnwidth,
height=0.95\textheight,
keepaspectratio
]{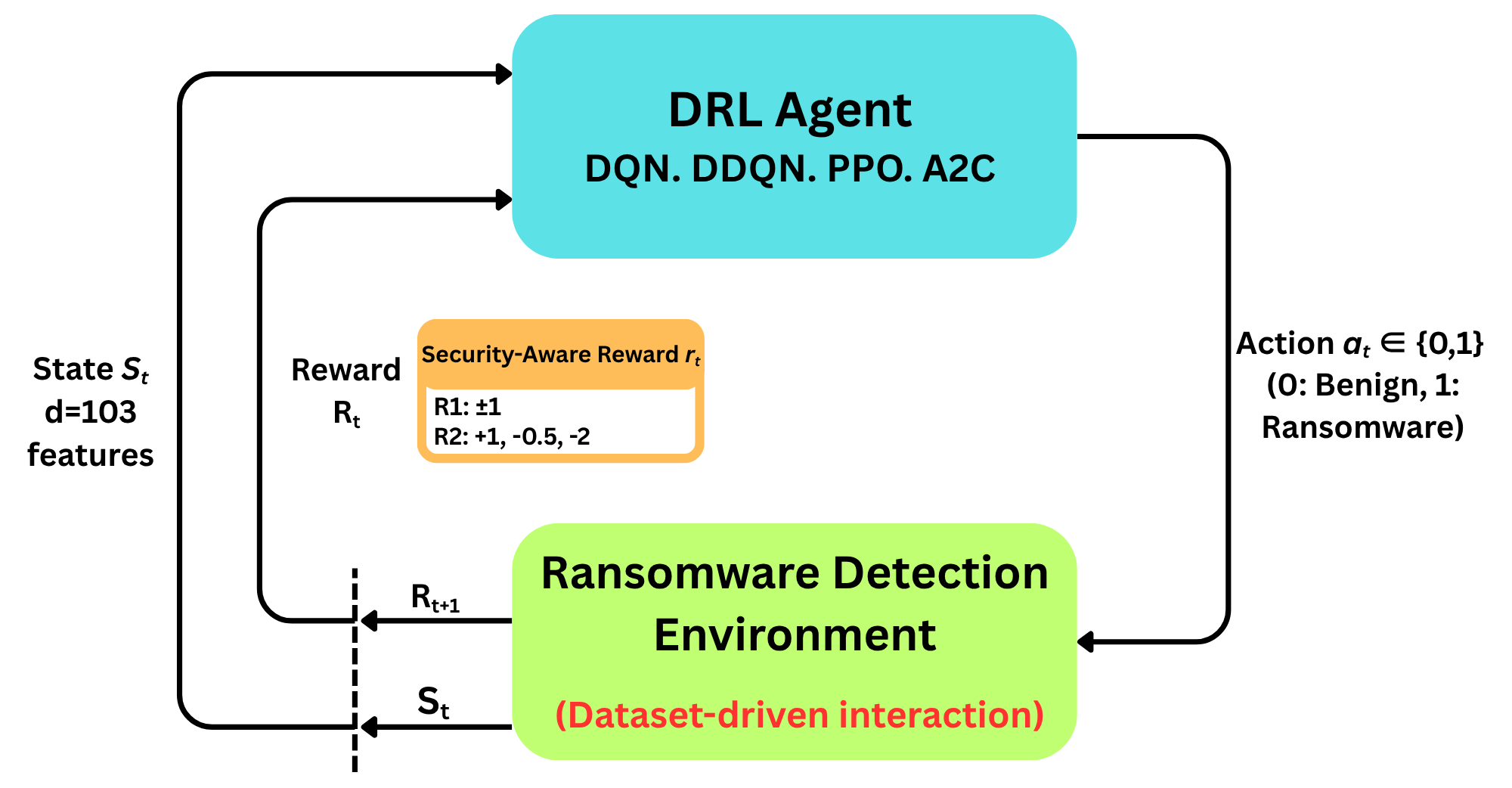}
\captionof{figure}{Security-aware deep reinforcement learning framework for ransomware detection. The agent learns optimal classification policies from behavioral state representations using asymmetric reward-driven interaction with the environment.}
\label{fig:drl_framework}
\end{center}

\subsubsection{Deep Q-Network (DQN)}
\label{subsubsec:dqn}

DQN extends traditional Q-learning by using a deep neural network to approximate action-value functions. Conventional Q-learning performs effectively in small finite environments but becomes impractical in high-dimensional state spaces because of the exponential growth of the Q-table. DQN addresses this limitation by learning a parameterized approximation of the action-value function $Q(s,a;\theta)$ using deep neural networks.

The DQN objective minimizes the temporal-difference (TD) loss:
\begin{equation}
\mathcal{L}(\theta)
=
\mathbb{E}
\left[
\left(
r_t
+
\gamma
\max_{a'}
Q(s_{t+1},a';\theta^{-})
-
Q(s_t,a_t;\theta)
\right)^2
\right]
\label{eq:dqn_loss}
\end{equation}
where $\theta^{-}$ denotes the target-network parameters. The agent selects actions using a greedy policy with respect to the learned Q-values.

\subsubsection{Double Deep Q-Network (DDQN)}
\label{subsubsec:ddqn}

Standard Q-learning and DQN use the same network for both action selection and value estimation. Consequently, noisy Q-value estimates may become systematically amplified, leading to overestimation bias and unstable policy learning.

DDQN mitigates this problem by decoupling action selection and action evaluation. The online network selects the best next action, while the target network evaluates its value:
\begin{equation}
Y_t
=
r_t
+
\gamma
Q
\left(
s_{t+1},
\arg\max_{a'}
Q(s_{t+1},a';\theta),
\theta^{-}
\right)
\label{eq:ddqn_target}
\end{equation}
This separation reduces optimistic bias and produces more stable and reliable Q-value estimates. Such stability is particularly important under asymmetric reward settings, where inaccurate value estimation may lead to costly missed ransomware detections.

\subsubsection{Proximal Policy Optimization (PPO)}
\label{subsubsec:ppo}

PPO is a policy-gradient method that directly optimizes the classification policy while constraining excessively large policy updates through a clipping mechanism. In the proposed framework, PPO learns the probability of assigning each behavioral state to either the ransomware or benign class.

The PPO objective is defined as:
\begin{equation}
\mathcal{L}^{\mathrm{PPO}}(\theta)
=
\mathbb{E}
\left[
\min
\left(
\rho_t(\theta)\hat{A}_t,
\mathrm{clip}
\left(
\rho_t(\theta),
1-\epsilon,
1+\epsilon
\right)
\hat{A}_t
\right)
\right]
\label{eq:ppo_loss}
\end{equation}
where
\begin{equation}
\rho_t(\theta)
=
\frac{
\pi_{\theta}(a_t|s_t)
}{
\pi_{\theta_{\mathrm{old}}}(a_t|s_t)
}
\label{eq:ppo_ratio}
\end{equation}
and $\hat{A}_t$ denotes the estimated advantage function. PPO improves optimization stability by restricting abrupt policy changes during training.

\subsubsection{Advantage Actor--Critic (A2C)}
\label{subsubsec:a2c}

A2C combines an actor network that learns the classification policy with a critic network that estimates the expected state value. The critic provides a learned baseline through the advantage function, reducing gradient variance and accelerating convergence.

The advantage estimate is defined as:
\begin{equation}
A_t
=
r_t
+
\gamma V(s_{t+1})
-
V(s_t)
\label{eq:a2c_advantage}
\end{equation}
where $V(s)$ denotes the learned state-value function. The actor updates the policy using the estimated advantage, while the critic simultaneously learns state-value estimation.

\subsubsection{Comparison and Suitability for Security-Aware Learning}
\label{subsubsec:model_comparison}

The evaluated DRL models differ fundamentally in how decision policies are learned. Value-based methods (DQN and DDQN) directly estimate action values, whereas policy-gradient methods (PPO and A2C) optimize action probabilities through policy parameterization.

Under the proposed asymmetric reward function $R_2$, accurate value estimation becomes particularly important because small estimation errors may lead to disproportionately costly false negatives. DDQN is especially suitable for this setting because it mitigates overestimation bias and produces more reliable decision boundaries under asymmetric reward landscapes. By contrast, policy-gradient methods rely on advantage estimation, which may introduce higher optimization variance under sharply asymmetric rewards, particularly at larger discount factors.

\subsubsection{Training Protocol}
\label{subsubsec:training_protocol}

All DRL agents were trained for 10,000 timesteps per experimental run. The complete factorial experimental design is defined as:
\begin{equation}
|\mathrm{Algos}|
\times
|\mathrm{Rewards}|
\times
|\Gamma|
\times
K
\times
|S|
=
4
\times
2
\times
4
\times
5
\times
3
=
480
\label{eq:factorial_design}
\end{equation}
where $|\mathrm{Algos}|$ denotes the number of DRL algorithms, $|\mathrm{Rewards}|$ denotes the number of reward functions, $|\Gamma|$ denotes the number of discount factors, $K$ denotes the number of cross-validation folds, and $|S|$ denotes the number of random seeds.

During evaluation, all agents used deterministic policy execution (\texttt{deterministic=True}). DQN and DDQN employed greedy action selection, whereas PPO and A2C selected the highest-probability policy action.

For AUC computation, continuous prediction scores were used instead of binary labels. Value-based agents used the Q-value margin:
\begin{equation}
Q(s,1;\theta)-Q(s,0;\theta)
\label{eq:q_margin}
\end{equation}
whereas policy-gradient agents used the predicted ransomware probability:
\begin{equation}
\pi_{\theta}(a=1|s)
\label{eq:policy_probability}
\end{equation}
These continuous outputs provide threshold-independent discrimination estimates for ROC-AUC analysis.

\subsection{Security-Optimal Model Selection (SOMS)}
\label{subsec:soms}

To identify the most suitable model for security-critical ransomware detection, this study introduces a hierarchical model-selection rule that prioritizes minimum false-negative rate (FNR), followed by maximum F1-score, and finally minimum training time. This ordering reflects operational ransomware-defense priorities, where missed attacks represent the most severe failure mode, while predictive balance and computational efficiency remain secondary but important considerations.

Formally, the proposed Security-Optimal Model Selection (SOMS) rule is defined as the following lexicographic optimization problem:
\begin{equation}
\mathcal{M}^{\ast}
=
\operatorname*{lexmin}_{\mathcal{M}\in\mathcal{H}}
\left(
\mathrm{FNR}(\mathcal{M}),
-
\mathrm{F1}(\mathcal{M}),
T_{\mathrm{train}}(\mathcal{M})
\right)
\label{eq:soms}
\end{equation}
where $\mathcal{H}$ denotes the set of candidate models, $\mathcal{M}$ denotes an individual model within $\mathcal{H}$, and $\mathcal{M}^{\ast}$ denotes the model selected as optimal under the SOMS criterion. Here, $\mathrm{FNR}(\mathcal{M})$ represents the false-negative rate of model $\mathcal{M}$, $\mathrm{F1}(\mathcal{M})$ denotes its F1-score, and $T_{\mathrm{train}}(\mathcal{M})$ denotes its training time.

The operator $\operatorname{lexmin}$ denotes lexicographic minimization. Models are first compared according to FNR; only when two or more models exhibit similar FNR values is the F1-score considered as the second criterion. If a tie remains, training time is used as the final tie-breaking criterion. The negative sign preceding $\mathrm{F1}(\mathcal{M})$ converts F1-score maximization into an equivalent minimization problem.

The SOMS criterion is directly aligned with the security-aware reward design presented in Section~\ref{subsec:reward_design}, where false negatives receive substantially larger penalties than false positives. While the reward function shapes policy learning during training, SOMS governs final model selection under deployment-oriented security constraints.

The rationale underlying this hierarchy is fundamentally security-driven. In ransomware detection, a false negative corresponds to an undetected malicious execution that may lead to file encryption, operational disruption, and financial loss. Consequently, minimizing FNR constitutes the primary optimization objective. Among models exhibiting comparable FNR values, the F1-score preserves predictive balance between precision and recall. Training time serves as a tertiary criterion that promotes computational efficiency without overriding the primary security objective.

Accordingly, SOMS transforms model selection from a generic performance-oriented comparison into a security-aware decision framework. This distinction is particularly important because different DRL agents may achieve similar aggregate accuracy while exhibiting materially different false-negative behaviors. By explicitly enforcing the priority structure:
\begin{equation}
\min(\mathrm{FNR})
\rightarrow
\max(\mathrm{F1})
\rightarrow
\min(T_{\mathrm{train}})
\label{eq:soms_priority}
\end{equation}
the proposed framework ensures that the selected model aligns with the operational requirements of high-risk ransomware detection by minimizing missed attacks while preserving predictive balance and computational efficiency.

Algorithm~\ref{alg:sadr_training} summarizes the complete SA-DRL training and SOMS-based model-selection workflow.

\begin{algorithm*}[t]
\caption{Security-Aware DRL Training and SOMS-Based Model Selection}
\label{alg:sadr_training}
\footnotesize

\KwIn{Dataset $\mathcal{D}=\{(x_i,y_i)\}_{i=1}^{N}$, $N=2000$, $d=103$; agents $\mathcal{M}=\{\mathrm{DQN,DDQN,PPO,A2C}\}$; rewards $\mathcal{R}=\{R_1,R_2\}$; discount factors $\Gamma=\{0.10,0.50,0.90,0.99\}$; folds $K=5$; seeds $S=\{42,123,456\}$; total training steps $T_{\mathrm{steps}}=10{,}000$.}

\KwOut{Security-optimal model $\mathcal{M}^{\ast}$ and learned policy $\pi^{\ast}$.}
\hrule
\vspace{2pt}
Collect 1,000 ransomware and 1,000 benign samples\;
Execute all samples in ANY.RUN under Windows~11\;
Extract behavioral JSON telemetry and construct $X\in\mathbb{R}^{2000\times103}$ and labels $y\in\{0,1\}^{2000}$\;
Initialize candidate set $\mathcal{H}\leftarrow\emptyset$\;

\ForEach{seed $s\in S$}{
Generate $K=5$ stratified train--test folds\;

\ForEach{fold $k=1$ to $K$}{
Split $\mathcal{D}$ into $\mathcal{D}_{\mathrm{train}}^{(k)}$ and $\mathcal{D}_{\mathrm{test}}^{(k)}$\;
Set $N_{\mathrm{train}}=|\mathcal{D}_{\mathrm{train}}^{(k)}|$\;
Fit preprocessing only on $\mathcal{D}_{\mathrm{train}}^{(k)}$, then transform the training and testing sets\;

\ForEach{reward $R\in\mathcal{R}$}{
\ForEach{agent $M\in\mathcal{M}$}{
\ForEach{discount factor $\gamma\in\Gamma$}{
Define MDP $\mathcal{E}=(\mathcal{S},\mathcal{A},P,R,\gamma)$, where $\mathcal{S}\subseteq\mathbb{R}^{103}$ and $\mathcal{A}=\{0,1\}$\;
Train $M$ for $T_{\mathrm{steps}}=10{,}000$ total timesteps\;
At each episode boundary, sample permutation $\pi:\{1,\ldots,N_{\mathrm{train}}\}\rightarrow\{1,\ldots,N_{\mathrm{train}}\}$\;

\ForEach{step $t$}{
Observe $s_t=x_{\pi(t)}$\;
Select action $a_t\in\{0,1\}$\;
Compute reward $r_t$\;
Transition to $s_{t+1}=x_{\pi(t+1)}$\;
}

Use reward functions\;
$R_1(a_t,y_t)=+1$ if $a_t=y_t$, otherwise $-1$\;
$R_2(a_t,y_t)=+1$ if $a_t=y_t$; $-2.0$ if $y_t=1$ and $a_t=0$; $-0.5$ if $y_t=0$ and $a_t=1$\;
Update $M$ using the corresponding DRL objective defined in Equations~(\ref{eq:dqn_loss})--(\ref{eq:a2c_advantage})\;
Evaluate $M$ on $\mathcal{D}_{\mathrm{test}}^{(k)}$ and record FNR, F1-score, and training time\;
Store configuration $c=(M,R,\gamma,k,s,\mathrm{FNR},\mathrm{F1},T_{\mathrm{train}},\pi_M)$ in $\mathcal{H}$\;
}
}
}
}
}

Select the security-optimal configuration using SOMS\;
$\mathcal{M}^{\ast}=\operatorname{lexmin}_{c\in\mathcal{H}}\left(\mathrm{FNR}(c),-\mathrm{F1}(c),T_{\mathrm{train}}(c)\right)$\;
Return $\mathcal{M}^{\ast}$ and learned policy $\pi^{\ast}$\;

\end{algorithm*}

\subsection{Baseline Models}
\label{subsec:baseline_models}

To contextualize the advantages of the proposed security-aware reward design, three supervised baseline models were evaluated alongside the DRL agents: a Multi-Layer Perceptron (MLP), a one-dimensional Convolutional Neural Network (CNN1D), and Logistic Regression (LR). The MLP represents nonlinear supervised classification through fully connected layers, CNN1D captures local feature dependencies across the behavioral feature sequence, and LR serves as a robust linear reference model.

All baseline models were trained and evaluated using the same 5-fold stratified cross-validation and three-seed replication protocol applied to the DRL agents. Identical preprocessing was enforced across all models, including StandardScaler normalization fitted exclusively on the training fold. In addition, all supervised baselines used the same security-sensitive class-weighting configuration, with SecurityCost weights defined as $w_{\mathrm{FN}}=5.0$ and $w_{\mathrm{FP}}=1.0$, ensuring that comparisons reflected genuine modeling differences rather than evaluation asymmetries. The complete hyperparameter specifications are provided in Section~\ref{sec:experimental_setup}.

\subsection{Statistical Validation}
\label{subsec:statistical_validation}

Following the protocol recommended by Dem\v{s}ar~\cite{ref52} for statistically sound multi-classifier comparison, a two-stage non-parametric statistical validation procedure was applied to the experimental results. Per-fold F1-scores were first aggregated across folds for each $(M,R,\gamma,s)$ configuration, after which global and pairwise significance tests were conducted.

\subsubsection{Friedman Test}

A Friedman test was performed to determine whether the observed performance differences among the reinforcement learning agents were statistically significant. The global null hypothesis $H_0$ assumes that all four DRL agents exhibit equivalent predictive performance. The Friedman statistic is defined as
\begin{equation}
\mathcal{X}_F^2=
\frac{12n}{k(k+1)}
\left[
\sum_{j=1}^{k}\bar{R}_j^2
-
\frac{k(k+1)^2}{4}
\right]
\label{eq:friedman}
\end{equation}
where $n$ denotes the number of paired experimental conditions, $k=4$ is the number of DRL agents, and $\bar{R}_j$ is the mean rank of agent $j$. Under the null hypothesis, the statistic approximately follows a chi-square distribution:
\[
\mathcal{X}_F^2 \sim \chi^2(k-1=3)
\]

\subsubsection{Wilcoxon Signed-Rank Test}

Following the Friedman test, pairwise Wilcoxon signed-rank tests were conducted for post-hoc analysis. If the Friedman test rejected $H_0$, all six pairwise classifier comparisons were evaluated using paired F1-score differences defined as
\begin{equation}
d_i=
\bar{F1}_{M_i,i}
-
\bar{F1}_{M_j,i}
\label{eq:wilcoxon_diff}
\end{equation}
where $d_i$ represents paired performance difference between classifiers $M_i$ and $M_j$ under experimental condition $i$.

A second complementary set of Wilcoxon signed-rank tests was performed specifically to evaluate the statistical superiority of the asymmetric reward function ($R_2$) over the symmetric baseline reward ($R_1$). For each DRL agent (DQN, DDQN, PPO, and A2C), paired false-negative-rate (FNR) observations were collected across all 12 configurations per agent (four discount factors $\times$ three random seeds), yielding $n=12$ paired differences for each algorithm.

The null hypothesis assumes that $R_1$ and $R_2$ produce equivalent FNR distributions, whereas the alternative hypothesis assumes that $R_2$ yields significantly lower FNR values. In addition, a pooled Wilcoxon test across all agents ($n=48$) was conducted to provide an omnibus assessment of the reward-function effect.

The results of all per-agent and pooled statistical tests are reported in Table~\ref{tab:wilcoxon_reward} in Section 5, providing statistical confirmation that the FNR reductions observed under $R_2$ are not attributable to stochastic training variance.

Statistical significance is reported using four thresholds:
$^{***}(p<0.001)$,
$^{**}(p<0.01)$,
$^{*}(p<0.05)$,
and $\mathrm{ns}(p\geq0.05)$.

\section{Experimental Setup}
\label{sec:experimental_setup}

\subsection{Dataset and Preprocessing}
\label{subsec:dataset_preprocessing}

All experiments were conducted using the balanced behavioral ransomware dataset described in Section~\ref{subsec:dataset}. Features were standardized using \texttt{StandardScaler} fitted exclusively on the training fold to prevent data leakage. The fitted transformation was subsequently applied to both the training and testing partitions.

\subsection{Cross-Validation and Replication Protocol}
\label{subsec:cv_protocol}

Model evaluation employed 5-fold stratified cross-validation using \texttt{StratifiedKFold} with
\texttt{n\_splits=5}, \texttt{shuffle=True}, and \texttt{random\_state=42}. This preserved the balanced 1:1 ransomware-to-benign class ratio across all folds, producing 1,600 training samples and 400 testing samples per fold.

To ensure reproducibility and reduce stochastic bias, all experimental configurations were replicated across three random seeds $\{42,123,456\}$, which were applied consistently to NumPy, PyTorch, and TensorFlow. All reported results are presented as mean $\pm$ standard deviation over 15 observations per configuration (5 folds $\times$ 3 seeds).

\subsection{Factorial Design}
\label{subsec:factorial_design}

The complete experiment followed a $4 \times 2 \times 4$ factorial design crossed with 5-fold cross-validation and 3-seed replication, resulting in a total of 480 DRL training runs.

\begin{table}[t]
\caption{Experimental factorial design.}
\label{tab:factorial_design}
\centering
\footnotesize
\begin{tabular}{lll}
\toprule
\textbf{Factor} & \textbf{Levels} & \textbf{Values} \\
\midrule
DRL algorithm &
4 &
DQN, DDQN, PPO, A2C \\

Reward function &
2 &
R1 (symmetric), R2 (asymmetric) \\

Discount factor $\gamma$ &
4 &
0.1, 0.5, 0.9, 0.99 \\

CV folds &
5 &
--- \\

Random seeds &
3 &
42, 123, 456 \\

Total DRL runs &
480 &
$4 \times 2 \times 4 \times 5 \times 3$ \\
\bottomrule
\end{tabular}
\end{table}

\subsection{DRL Hyperparameter Configurations}
\label{subsec:drl_hyperparameters}

Table~\ref{tab:drl_hyperparameters} summarizes all DRL hyperparameters implemented in the experimental framework.

\begin{table*}[t]
\caption{DRL agent hyperparameter configuration.}
\label{tab:drl_hyperparameters}
\centering
\footnotesize
\begin{tabular}{lllll}
\toprule
\textbf{Hyperparameter} &
\textbf{DQN} &
\textbf{DDQN} &
\textbf{PPO} &
\textbf{A2C} \\
\midrule

Network architecture &
$2 \times [64]$ ReLU &
$2 \times [64]$ ReLU &
$2 \times [64]$ ReLU &
$2 \times [64]$ ReLU \\

Learning rate $(\alpha)$ &
$1\times10^{-3}$ &
$1\times10^{-3}$ &
$3\times10^{-4}$ &
$7\times10^{-4}$ \\

Replay buffer size &
10,000 &
10,000 &
--- &
--- \\

Mini-batch size &
32 &
32 &
64 &
--- \\

Rollout steps $(n_{\mathrm{steps}})$ &
--- &
--- &
2,048 &
5 \\

Target update interval &
default &
500 &
--- &
--- \\

Discount factor $(\gamma)$ &
$\{0.1,0.5,0.9,0.99\}$ &
$\{0.1,0.5,0.9,0.99\}$ &
$\{0.1,0.5,0.9,0.99\}$ &
$\{0.1,0.5,0.9,0.99\}$ \\

Total timesteps &
10,000 &
10,000 &
10,000 &
10,000 \\

Policy &
MlpPolicy &
MlpPolicy &
MlpPolicy &
MlpPolicy \\
\bottomrule
\end{tabular}
\end{table*}

\noindent
\textit{Notes:}
``---'' denotes not applicable. Replay buffer and mini-batch parameters apply only to off-policy methods (DQN and DDQN). Rollout steps apply exclusively to on-policy methods (PPO and A2C). A2C performs a single gradient update over the complete $n_{\mathrm{steps}}=5$ rollout without mini-batch subdivision.

\subsection{Evaluation Metrics}
\label{subsec:evaluation_metrics}
\noindent

The performance of each configuration was evaluated using a combination of classification, security, and computational-efficiency metrics. Standard classification metrics quantify predictive performance, whereas false-negative and false-positive rates explicitly capture operational security risk. In addition, training and inference times were recorded to assess computational feasibility and deployment practicality.

Nine evaluation metrics were computed identically across all DRL and baseline models using the same \texttt{compute\_metrics()} implementation to ensure evaluation consistency. The evaluation metrics used in this study are summarized in Table~\ref{tab:evaluation_metrics}.

\begin{table*}[t]
\caption{Evaluation metrics were used to assess the predictive performance, security risk, and computational efficiency of the proposed ransomware detection framework.}
\label{tab:evaluation_metrics}
\centering
\small
\renewcommand{\arraystretch}{1.15}
\setlength{\tabcolsep}{5pt}

\begin{tabular*}{\textwidth}{@{\extracolsep{\fill}} p{2.9cm} p{5.8cm} p{4.4cm} p{3.3cm}}
\toprule

\textbf{Metric Name} &
\textbf{Description} &
\textbf{Related Syntax/Equation} &
\textbf{Security Relevance} \\

\midrule

Accuracy &
The ratio of correctly predicted instances (TP and TN) to the total number of observations. &
$Accuracy=\frac{TP+TN}{TP+TN+FP+FN}$ &
General performance indicators. \\

Precision &
The ratio of correctly predicted positive observations (TP) to the total predicted positives (TP + FP). &
$Precision=\frac{TP}{TP+FP}$ &
False alarm cost \\

Recall (Sensitivity) &
It measures the proportion of actual ransomware samples (TP) correctly identified by the model. &
$Recall=\frac{TP}{TP+FN}$ &
Detection completeness \\

F1-Score &
The harmonic mean of precision and recall balances false positives and false negatives. &
$F1\text{-}measure=
2\times
\frac{Precision\times Recall}
{Precision+Recall}$ &
Harmonic balance of Prec and Rec \\

False Positive Rate (FPR) &
It measures the proportion of benign samples incorrectly classified as ransomware. &
$FPR=\frac{FP}{FP+TN}$ &
Primary: missed ransomware rate \\

False Negative Rate (FNR) &
It measures the proportion of ransomware samples incorrectly classified as benign. &
$FNR=\frac{FN}{FN+TP}$ &
Secondary: benign false alarm rate \\

Training Time &
Time required to train each RL model. &
Measured in seconds &
Deployment feasibility \\

Inference Time &
Time required to generate predictions from the test data. &
Measured in seconds &
Real-time applicability. \\

\bottomrule
\end{tabular*}
\end{table*}

\subsection{Implementation Environment}
\label{subsec:implementation_environment}

All experiments were conducted in a reproducible CPU-only cloud environment to reflect practical deployment constraints and computational accessibility. Table~\ref{tab:implementation_environment} summarizes the complete hardware, software, and implementation configuration used throughout the study.

\begin{table}[H]
\caption{Implementation environment and software configuration used for experimental evaluation.}
\label{tab:implementation_environment}
\centering
\footnotesize
\begin{tabular}{ll}
\toprule
\textbf{Item} & \textbf{Specification} \\
\midrule

Behavioral analysis platform &
Windows 11 ANY.RUN sandbox \\

Execution environment &
Google Colab Pro (CPU-only) \\

Available RAM &
51 GB \\

Baseline models &
MLP, LR, CNN1D \\

Statistical tests &
Friedman, Wilcoxon \\

DRL backend &
PyTorch 2.1 \\

Programming language &
Python 3.10 \\

Preprocessing and validation &
scikit-learn 1.3 \\

Visualization libraries &
matplotlib, seaborn \\

\bottomrule
\end{tabular}
\end{table}


\section{Experimental Results Analysis}
\label{sec:results}

\subsection{Overview of Results}
\label{subsec:overview_results}

This section evaluates the proposed SA-DRL framework in all experimental configurations (Tables~\ref{tab:seed42_results}--\ref{tab:seed456_results}) to analyze the effects of reward design, DRL architecture and discount-factor selection on the security-oriented ransomware detection performance. In general, value-based methods (DDQN and DQN) demonstrate stronger ransomware detection capabilities than policy-gradient methods (PPO and A2C), particularly in terms of lower false-negative rates. The results also indicate that the reward function $R_2$ generally produces more security-oriented behavior than the baseline design $R_1$ by reducing the number of missed ransomware detections. Additionally, smaller discount factors, especially $\gamma=0.10$ and $\gamma=0.50$, are more frequently associated with strong and stable performance, whereas larger values such as $\gamma=0.90$ and $\gamma=0.99$ tend to degrade security-oriented performance, particularly for PPO and A2C.

Taken together, Tables~\ref{tab:seed42_results}--\ref{tab:seed456_results} provide consistent evidence that favors DDQN/DQN over PPO/A2C, $R_2$ over $R_1$ for security-oriented detection, and lower discount factors as the most suitable choice for this task. Across all three random seeds, the reward function $R_2$ consistently reduced the FNR of DDQN and DQN relative to $R_1$, with the strongest improvements observed under lower discount factors ($\gamma=0.10$ and $\gamma=0.50$). Importantly, these reductions occurred without substantial degradation in overall F1-score, confirming that asymmetric reward shaping successfully biases the learning objective toward ransomware-miss avoidance while preserving balanced predictive performance.

Complete per-configuration experimental results across all algorithms, reward functions, discount factors, and random
seeds are provided in the Supplementary Material.

Given the observed variation across DRL agents and reward configurations, the proposed SOMS framework was subsequently applied to identify the most security-optimal deployment configuration.

\begin{table*}[!t]
\caption{Best configuration for each algorithm--reward pair under Seed 42, selected using the security-aware comparison rule. Complete per-configuration results are provided in the Supplementary Material.}
\label{tab:seed42_results}

\centering
\scriptsize
\setlength{\tabcolsep}{3pt}

\resizebox{\textwidth}{!}{%

\begin{tabular}{lllllllllll}
\toprule

\textbf{Alg.} &
\textbf{Reward} &
\textbf{$\gamma$} &
\textbf{Acc.} &
\textbf{Prec.} &
\textbf{Recall} &
\textbf{F1} &
\textbf{FNR} &
\textbf{FPR} &
\textbf{Train (s)} &
\textbf{Infer (s)} \\

\midrule

DDQN & R1 & 0.10 &
$0.9900\pm0.0040$ &
$0.9930\pm0.0028$ &
$0.9870\pm0.0057$ &
$0.9900\pm0.0040$ &
$0.0130\pm0.0057$ &
$0.0070\pm0.0027$ &
$10.2825\pm0.9855$ &
$0.0848\pm0.0072$ \\

DDQN & R2 & 0.10 &
$0.9900\pm0.0031$ &
$0.9900\pm0.0049$ &
$0.9900\pm0.0061$ &
$0.9900\pm0.0031$ &
$0.0100\pm0.0061$ &
$0.0100\pm0.0050$ &
$9.7517\pm0.4779$ &
$0.1281\pm0.0069$ \\

DQN & R1 & 0.50 &
$0.9920\pm0.0021$ &
$0.9960\pm0.0042$ &
$0.9880\pm0.0045$ &
$0.9920\pm0.0021$ &
$0.0120\pm0.0045$ &
$0.0040\pm0.0042$ &
$10.1439\pm0.3361$ &
$0.0810\pm0.0024$ \\

DQN & R2 & 0.50 &
$0.9910\pm0.0045$ &
$0.9920\pm0.0057$ &
$0.9900\pm0.0071$ &
$0.9910\pm0.0046$ &
$0.0100\pm0.0071$ &
$0.0080\pm0.0057$ &
$10.0447\pm0.2447$ &
$0.0787\pm0.0027$ \\

PPO & R1 & 0.10 &
$0.9810\pm0.0065$ &
$0.9919\pm0.0103$ &
$0.9700\pm0.0106$ &
$0.9808\pm0.0066$ &
$0.0300\pm0.0106$ &
$0.0080\pm0.0104$ &
$13.9314\pm0.2375$ &
$0.1627\pm0.0050$ \\

PPO & R2 & 0.90 &
$0.9830\pm0.0021$ &
$0.9820\pm0.0027$ &
$0.9840\pm0.0042$ &
$0.9830\pm0.0021$ &
$0.0160\pm0.0042$ &
$0.0180\pm0.0027$ &
$13.6602\pm0.2782$ &
$0.1617\pm0.0091$ \\

A2C & R1 & 0.50 &
$0.9800\pm0.0061$ &
$0.9898\pm0.0036$ &
$0.9700\pm0.0106$ &
$0.9798\pm0.0063$ &
$0.0300\pm0.0106$ &
$0.0100\pm0.0035$ &
$16.1046\pm0.1864$ &
$0.1616\pm0.0073$ \\

A2C & R2 & 0.10 &
$0.9765\pm0.0070$ &
$0.9838\pm0.0089$ &
$0.9690\pm0.0096$ &
$0.9763\pm0.0070$ &
$0.0310\pm0.0096$ &
$0.0160\pm0.0089$ &
$16.0965\pm0.4827$ &
$0.2952\pm0.0063$ \\

\bottomrule
\end{tabular}

}
\end{table*}


\begin{table*}[!t]
\caption{Best configuration for each algorithm--reward pair under Seed 123, selected using the security-aware comparison rule. Complete per-configuration results are provided in the Supplementary Material.}
\label{tab:seed123_results}

\centering
\scriptsize
\setlength{\tabcolsep}{3pt}

\resizebox{\textwidth}{!}{%

\begin{tabular}{lllllllllll}
\toprule

\textbf{Alg.} &
\textbf{Reward} &
\textbf{$\gamma$} &
\textbf{Acc.} &
\textbf{Prec.} &
\textbf{Recall} &
\textbf{F1} &
\textbf{FNR} &
\textbf{FPR} &
\textbf{Train (s)} &
\textbf{Infer (s)} \\

\midrule

DDQN & R1 & 0.10 &
$0.9905\pm0.0021$ &
$0.9940\pm0.0042$ &
$0.9870\pm0.0027$ &
$0.9905\pm0.0021$ &
$0.0130\pm0.0027$ &
$0.0060\pm0.0042$ &
$10.2880\pm0.7089$ &
$0.0857\pm0.0093$ \\

DDQN & R2 & 0.10 &
$0.9875\pm0.0018$ &
$0.9832\pm0.0072$ &
$0.9920\pm0.0057$ &
$0.9876\pm0.0017$ &
$0.0080\pm0.0057$ &
$0.0170\pm0.0076$ &
$9.8675\pm0.2201$ &
$0.1335\pm0.0074$ \\

DQN & R1 & 0.10 &
$0.9900\pm0.0018$ &
$0.9910\pm0.0041$ &
$0.9890\pm0.0022$ &
$0.9900\pm0.0018$ &
$0.0110\pm0.0022$ &
$0.0090\pm0.0042$ &
$10.5649\pm0.5801$ &
$0.0901\pm0.0106$ \\

DQN & R2 & 0.99 &
$0.9905\pm0.0037$ &
$0.9901\pm0.0078$ &
$0.9910\pm0.0055$ &
$0.9905\pm0.0037$ &
$0.0090\pm0.0055$ &
$0.0100\pm0.0079$ &
$9.7441\pm0.2681$ &
$0.0800\pm0.0010$ \\

PPO & R1 & 0.50 &
$0.9810\pm0.0042$ &
$0.9898\pm0.0036$ &
$0.9720\pm0.0045$ &
$0.9808\pm0.0055$ &
$0.0180\pm0.0045$ &
$0.0170\pm0.0104$ &
$13.7766\pm0.2670$ &
$0.1651\pm0.0057$ \\

PPO & R2 & 0.50 &
$0.9825\pm0.0056$ &
$0.9831\pm0.0102$ &
$0.9820\pm0.0082$ &
$0.9825\pm0.0055$ &
$0.0180\pm0.0082$ &
$0.0170\pm0.0104$ &
$13.7766\pm0.2670$ &
$0.1621\pm0.0037$ \\

A2C & R1 & 0.50 &
$0.9770\pm0.0082$ &
$0.9780\pm0.0090$ &
$0.9760\pm0.0082$ &
$0.9770\pm0.0082$ &
$0.0240\pm0.0082$ &
$0.0220\pm0.0091$ &
$15.9409\pm0.3881$ &
$0.1555\pm0.0037$ \\

A2C & R2 & 0.10 &
$0.9825\pm0.0043$ &
$0.9908\pm0.0043$ &
$0.9740\pm0.0055$ &
$0.9823\pm0.0044$ &
$0.0260\pm0.0055$ &
$0.0090\pm0.0042$ &
$16.0717\pm0.4991$ &
$0.2934\pm0.0077$ \\

\bottomrule
\end{tabular}

}
\end{table*}


\begin{table*}[!t]
\caption{Best configuration for each algorithm--reward pair under Seed 456, selected using the security-aware comparison rule. Complete per-configuration results are provided in the Supplementary Material.}
\label{tab:seed456_results}

\centering
\scriptsize
\setlength{\tabcolsep}{3pt}

\resizebox{\textwidth}{!}{%

\begin{tabular}{lllllllllll}
\toprule

\textbf{Alg.} &
\textbf{Reward} &
\textbf{$\gamma$} &
\textbf{Acc.} &
\textbf{Prec.} &
\textbf{Recall} &
\textbf{F1} &
\textbf{FNR} &
\textbf{FPR} &
\textbf{Train (s)} &
\textbf{Infer (s)} \\

\midrule

DDQN & R1 & 0.10 &
$0.9900\pm0.0040$ &
$0.9930\pm0.0028$ &
$0.9870\pm0.0057$ &
$0.9900\pm0.0040$ &
$0.0130\pm0.0057$ &
$0.0070\pm0.0027$ &
$10.2825\pm0.9855$ &
$0.0848\pm0.0072$ \\

DDQN & R2 & 0.10 &
$0.9915\pm0.0038$ &
$0.9910\pm0.0041$ &
$0.9920\pm0.0057$ &
$0.9915\pm0.0038$ &
$0.0080\pm0.0057$ &
$0.0090\pm0.0042$ &
$9.9597\pm0.1799$ &
$0.1316\pm0.0015$ \\

DQN & R1 & 0.90 &
$0.9920\pm0.0021$ &
$0.9970\pm0.0028$ &
$0.9870\pm0.0027$ &
$0.9920\pm0.0021$ &
$0.0130\pm0.0027$ &
$0.0030\pm0.0028$ &
$10.6172\pm1.0102$ &
$0.0871\pm0.0102$ \\

DQN & R2 & 0.50 &
$0.9895\pm0.0033$ &
$0.9871\pm0.0045$ &
$0.9920\pm0.0057$ &
$0.9895\pm0.0033$ &
$0.0080\pm0.0057$ &
$0.0130\pm0.0045$ &
$10.2910\pm0.4132$ &
$0.0797\pm0.0015$ \\

PPO & R1 & 0.50 &
$0.9825\pm0.0053$ &
$0.9909\pm0.0056$ &
$0.9740\pm0.0096$ &
$0.9823\pm0.0054$ &
$0.0260\pm0.0096$ &
$0.0090\pm0.0057$ &
$14.3758\pm0.7314$ &
$0.1620\pm0.0090$ \\

PPO & R2 & 0.90 &
$0.9760\pm0.0014$ &
$0.9732\pm0.0054$ &
$0.9790\pm0.0055$ &
$0.9761\pm0.0014$ &
$0.0210\pm0.0055$ &
$0.0270\pm0.0057$ &
$13.8222\pm0.3553$ &
$0.1620\pm0.0097$ \\

A2C & R1 & 0.10 &
$0.9795\pm0.0057$ &
$0.9849\pm0.0079$ &
$0.9740\pm0.0074$ &
$0.9794\pm0.0057$ &
$0.0260\pm0.0074$ &
$0.0150\pm0.0079$ &
$16.0047\pm0.4379$ &
$0.1512\pm0.0035$ \\

A2C & R2 & 0.50 &
$0.9790\pm0.0068$ &
$0.9800\pm0.0110$ &
$0.9780\pm0.0027$ &
$0.9790\pm0.0068$ &
$0.0220\pm0.0027$ &
$0.0200\pm0.0112$ &
$16.1994\pm0.5593$ &
$0.1607\pm0.0034$ \\

\bottomrule
\end{tabular}

}
\end{table*}

\subsection{Security-Optimal Model Selection (SOMS)}
\label{subsec:soms_results}

The proposed Security-Optimal Model Selection (SOMS) framework was applied to identify the most suitable model under security-critical constraints. SOMS employs a lexicographic decision rule that prioritizes (i) the minimum false-negative rate (FNR), followed by (ii) the maximum F1-score, and finally (iii) the minimum training time. This reflects the operational reality that a missed ransomware execution is more costly than any other detection error.

Under the security-optimal configuration (Seed 456, $R_2$ reward, $\gamma=0.1$), Table~\ref{tab:soms_comparison} presents a full performance comparison of all four DRL agents and three supervised baselines. DDQN achieved the lowest FNR (0.0080), outperforming DQN (0.0090), PPO (0.0230), A2C (0.0280), and all baselines (MLP: 0.0247, LR: 0.0250, CNN1D: 0.0760). Relative to the best-performing supervised baseline on FNR (MLP, FNR = 0.0247), DDQN reduced missed ransomware detections by 67.6\% and outperformed all supervised baselines across FNR, FPR, and F1-score under identical evaluation conditions.


\begin{table*}[t]
\caption{Performance comparison of reinforcement learning algorithms under the selected security-oriented setting (Seed 456, $R_2$ reward function, $\gamma = 0.1$). DDQN achieved the lowest false-negative rate, strongest F1 performance, and lowest training and inference times, making it the preferred model under the security-aware comparison rule.}
\label{tab:soms_comparison}

\centering
\footnotesize
\setlength{\tabcolsep}{5pt}

\resizebox{\textwidth}{!}{%

\begin{tabular}{lcccccccc}
\toprule

\textbf{Algorithm} &
\textbf{Accuracy} &
\textbf{Precision} &
\textbf{Recall} &
\textbf{F1 Score} &
\textbf{FNR} &
\textbf{FPR} &
\textbf{Training Time (s)} &
\textbf{Inference Time (s)} \\

\midrule

DDQN &
0.9915 &
0.9910 &
0.9920 &
0.9915 &
0.0080 &
0.0090 &
9.96 &
0.1316 \\

DQN &
0.9905 &
0.9900 &
0.9910 &
0.9905 &
0.0090 &
0.0100 &
10.00 &
0.1323 \\

PPO &
0.9835 &
0.9899 &
0.9770 &
0.9834 &
0.0230 &
0.0100 &
13.83 &
0.2997 \\

A2C &
0.9795 &
0.9869 &
0.9720 &
0.9794 &
0.0280 &
0.0130 &
16.04 &
0.2982 \\

MLP &
0.9817 &
0.9879 &
0.9753 &
0.9815 &
0.0247 &
0.0120 &
0.9835 &
0.0018 \\

Logistic Regression (LR) &
0.9845 &
0.9939 &
0.9750 &
0.9843 &
0.0250 &
0.0060 &
0.0419 &
0.0005 \\

CNN1D &
0.9362 &
0.9479 &
0.9240 &
0.9354 &
0.0760 &
0.0517 &
12.2507 &
0.2667 \\

\bottomrule
\end{tabular}

}
\end{table*}

The confusion matrix for the SOMS-selected model is shown in Fig.~\ref{fig:conf_matrix}. Across 2,000 out-of-fold test samples, DDQN missed only eight ransomware samples (FN = 8, FNR = 0.80\%) while incorrectly flagging nine benign files (FP = 9, FPR = 0.90\%), yielding an accuracy of 99.15\% and an F1-score of 0.9915.

\begin{center}
 \centering
\includegraphics[width=0.98\columnwidth]{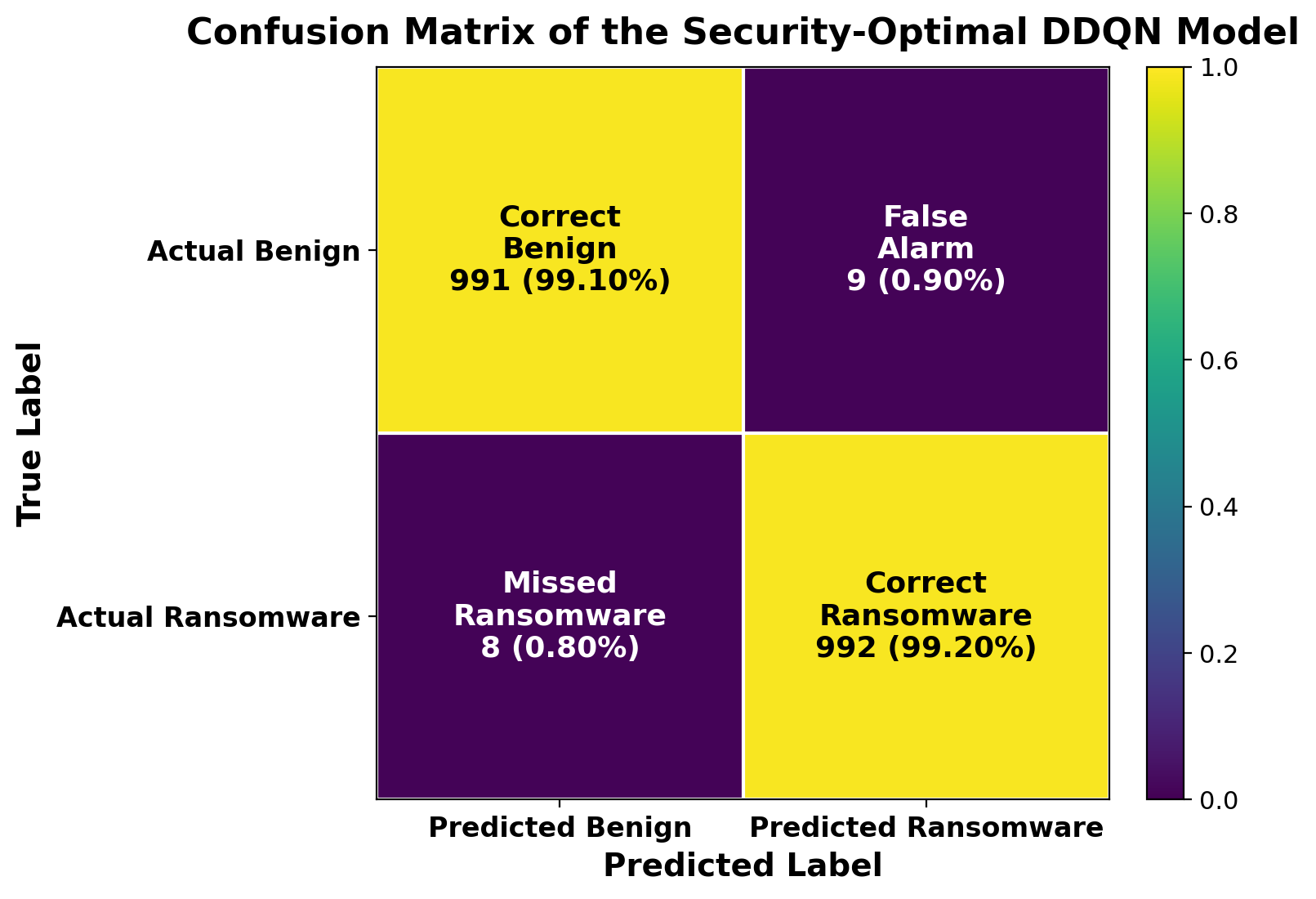}
\captionof{figure}{Confusion matrix of the security-optimal DDQN model under the $R_2$ reward setting and $\gamma = 0.1$. The model achieves a 0.8\% false-negative rate, minimizing missed ransomware detections, and a 0.9\% false-positive rate, demonstrating balanced performance.}
\label{fig:conf_matrix}
\end{center}

The score-based ROC curves for all four DRL agents under the same configuration ($R_2$, $\gamma = 0.1$) are shown in Fig.~\ref{fig:roc_curves}. DDQN achieved the highest AUC (0.998), followed by DQN (0.996), PPO (0.991), and A2C (0.983), confirming that the performance advantage of DDQN is threshold-independent and not an artifact of a specific classification threshold or operating point.

\begin{center}
\includegraphics[width=0.85\columnwidth]{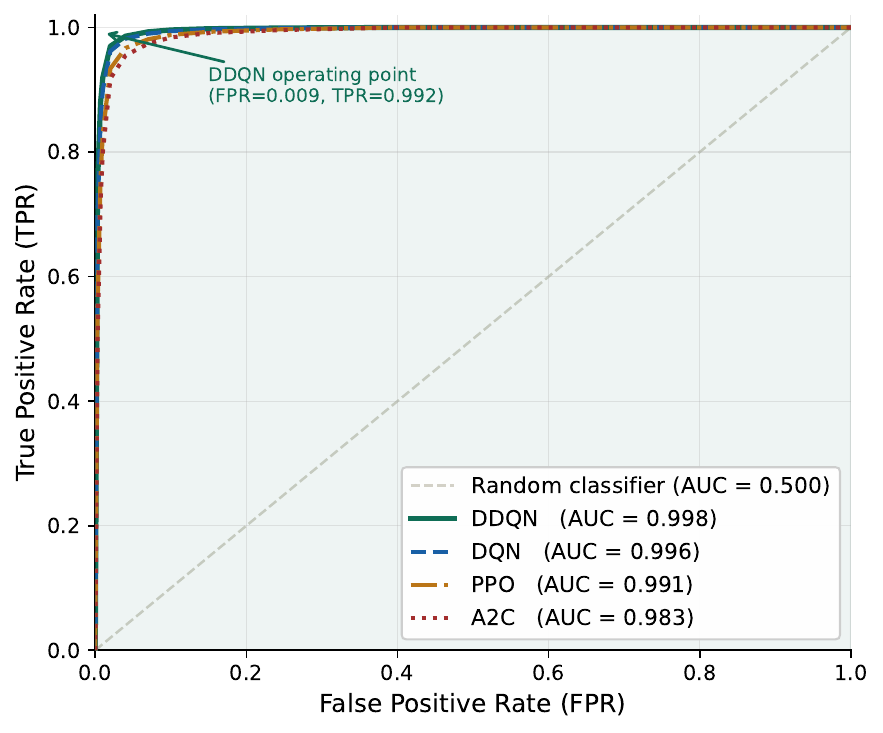}
\captionof{figure}{Score-based ROC curves for four DRL agents ($R_2$, $\gamma = 0.1$, Seed 456). DDQN has the highest AUC (0.998), with its joint FPR = 0.009 and TPR = 0.992, confirming threshold-independent superiority over DQN (0.996), PPO (0.991), and A2C (0.983).}
\label{fig:roc_curves}
\end{center}

Fig.~\ref{fig:pareto_frontier} presents the performance--efficiency trade-off for all seven evaluated models, plotting the F1-score against training time with the Pareto frontier overlaid. DDQN lies on the Pareto frontier as the SOMS winner; it achieves the highest F1-score among non-dominated models at a training cost of 9.96 s per fold, substantially below PPO (13.83 s) and A2C (16.04 s). Although LR offers the fastest training time (0.04 s), its FNR of 0.0250 renders it unsuitable for security-critical deployment.
Fig.~\ref{fig:time_comparison} further illustrates the training and inference time comparison on a logarithmic scale across all models, confirming that DDQN provides the best balance between detection performance and computational overhead among the DRL agents.

\begin{center}
\includegraphics[width=0.90\columnwidth]{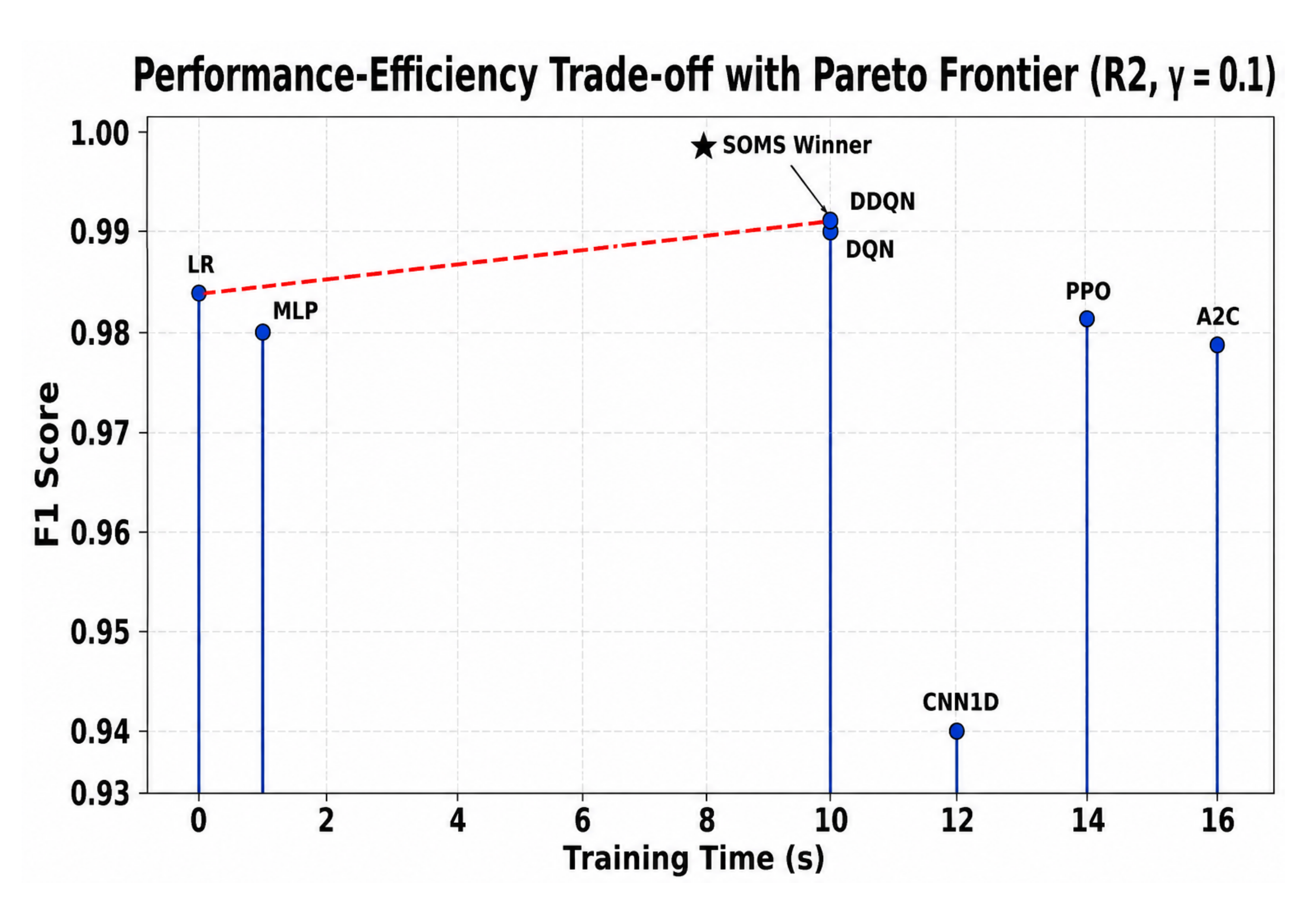}
\captionof{figure}{Performance--efficiency trade-off under the $R_2$ reward ($\gamma=0.1$). The Pareto frontier shows non-dominated models, and SOMS selects DDQN as the optimal model based on minimum FNR, maximum F1-score, and efficient training time.}
\label{fig:pareto_frontier}
\end{center}

\begin{figure*}[!t]
\centering
\includegraphics[width=0.80\textwidth]{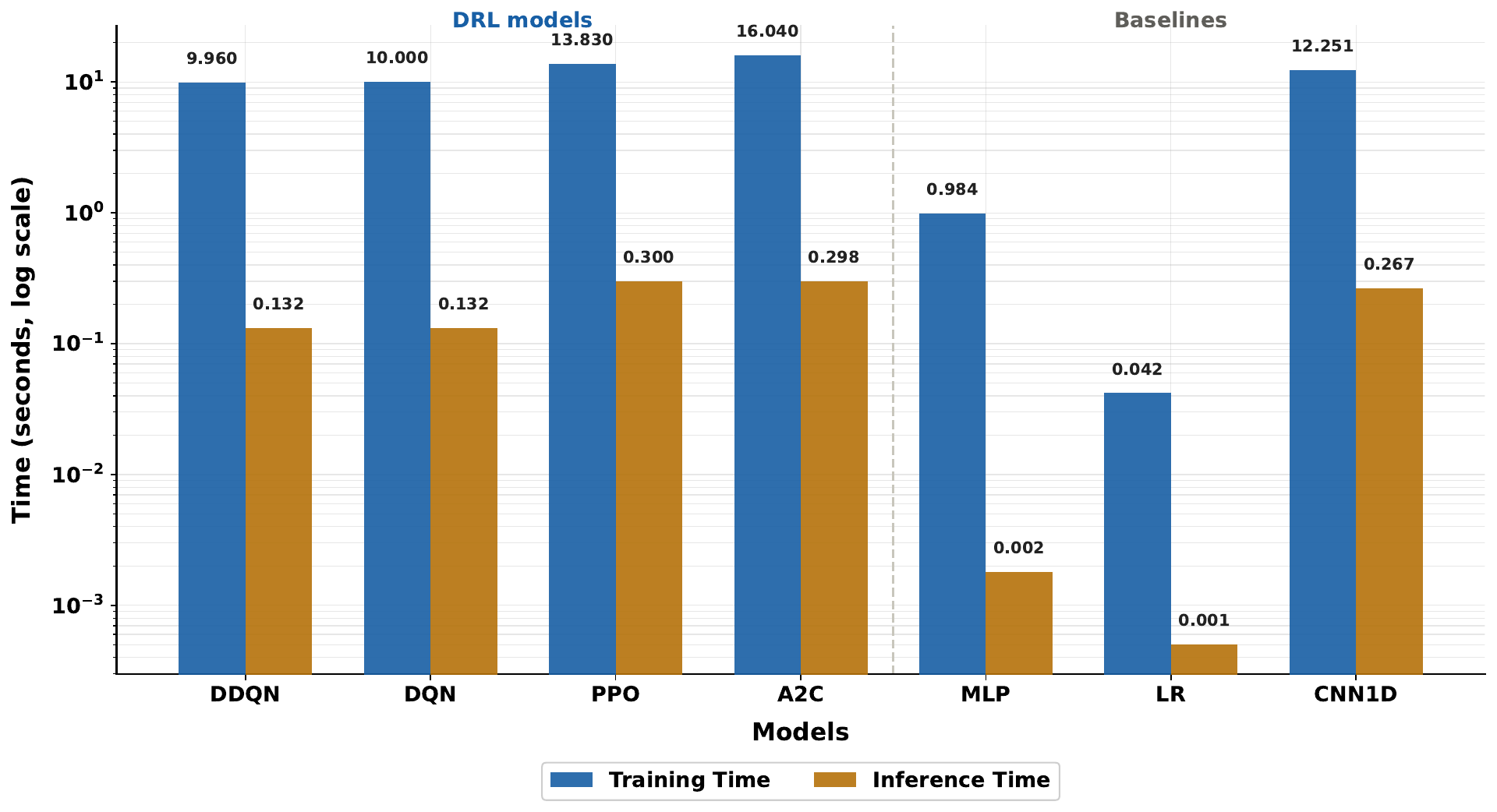}
\caption{Training and inference time comparison on a logarithmic scale across all DRL agents and supervised baselines ($R_2$, $\gamma=0.1$). DRL models require 9.96--16.04 s training time, whereas LR and MLP are substantially faster but exhibit higher FNR, limiting their suitability for security-critical deployment.}
\label{fig:time_comparison}
\end{figure*}

Although LR and MLP train and infer faster, their comparatively higher FNR limits their suitability for security-critical ransomware defense deployments. CNN1D is the weakest model (FNR = 0.0760) because one-dimensional convolution over an unordered behavioral feature vector is less effective at capturing the discriminative relationships exploited by tabular-learning and RL-based approaches. The superior performance of DDQN and DQN suggests that value-based temporal-difference optimization is better suited to security-sensitive binary classification under asymmetric reward conditions. Because the $R_2$ reward explicitly penalizes false negatives more heavily, stable Q-value estimation enables DDQN and DQN to propagate security-sensitive cost information more consistently across sequential updates. In contrast, PPO and A2C exhibited higher FNR variability, likely due to policy-gradient variance, less stable reward attribution, and sensitivity to asymmetric penalties under relatively small episodic datasets.

\subsection{Effect of Reward Function}
\label{subsec:reward_effect}

Figure~7 compares the mean FNR of DDQN, DQN, PPO, and A2C under the symmetric reward function ($R_1$) and the security-aware asymmetric reward function ($R_2$), averaged across all seeds and $\gamma$ values using the best configurations in Tables~1--3. The blue bars represent $R_1$, whereas the orange bars represent $R_2$. Across all agents, $R_2$ consistently produces lower FNR values than $R_1$, indicating that the asymmetric reward formulation improves ransomware detection by reducing missed detections. The most substantial reductions are observed for DDQN and PPO, both achieving a 35\% decrease in FNR, followed by DQN with a 25\% reduction. In contrast, A2C shows only a marginal improvement of approximately 1\%, suggesting weaker sensitivity to the reward redesign.

Among all agents, DDQN and DQN achieve the lowest overall FNR values under $R_2$, demonstrating stronger security-oriented detection capability. PPO exhibits the largest absolute reduction in FNR, decreasing from approximately 0.028 to 0.018, indicating that the asymmetric penalty structure strongly influences its policy optimization behavior. A2C remains comparatively unstable, with nearly identical FNR values under both reward formulations. Overall, the figure demonstrates that the asymmetric reward design effectively biases DRL agents toward reducing false negatives, which is critical for security-sensitive ransomware detection.

\begin{center}
\includegraphics[width=0.98\columnwidth]{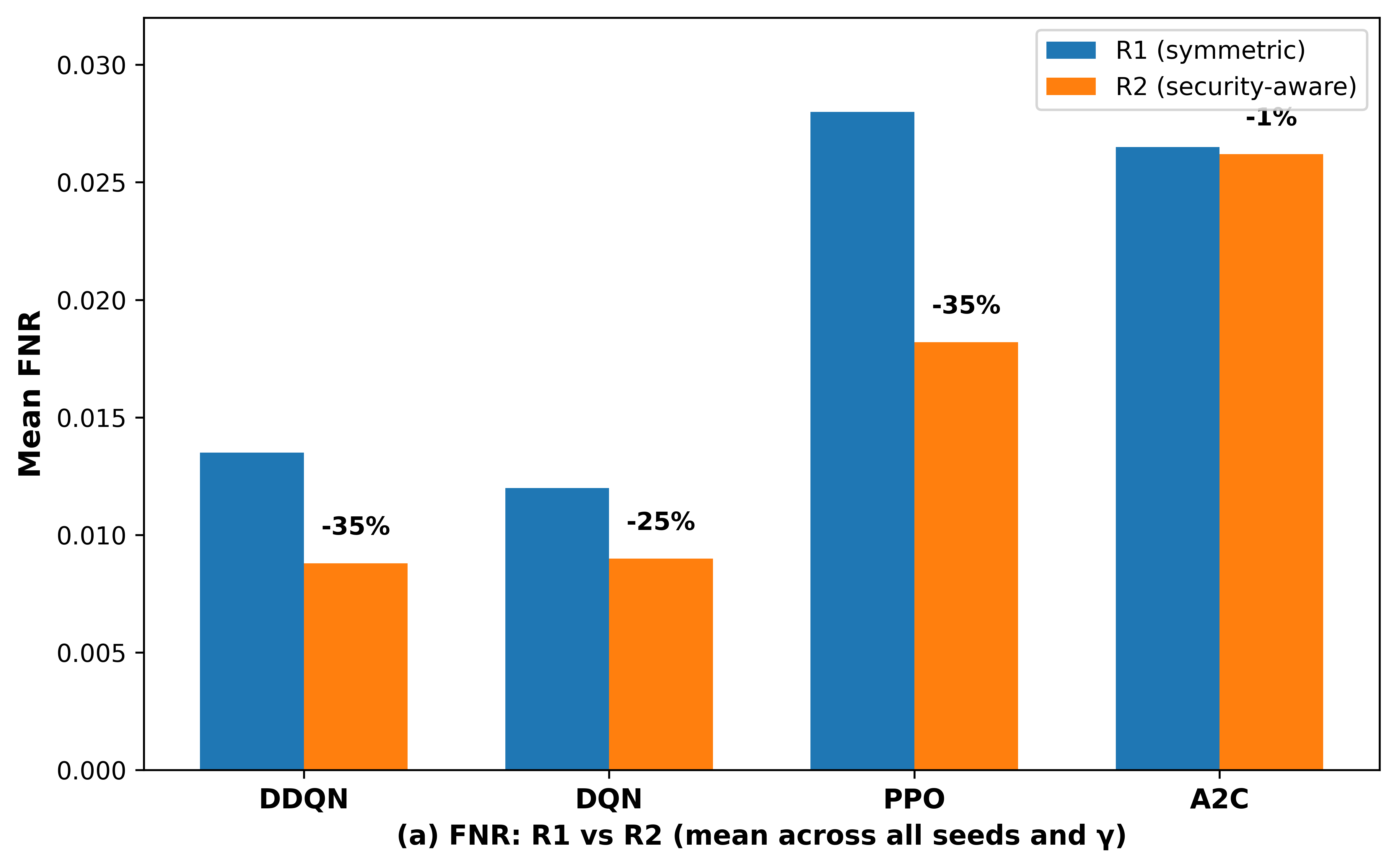}
\captionof{figure}{Comparison of mean FNR under $R_1$ (symmetric) and $R_2$ (security-aware) reward functions, averaged across seeds and discount factors. $R_2$ reduces FNR by 35\% for DDQN, 25\% for DQN, 35\% for PPO, and 1\% for A2C, confirming asymmetric reward design as the dominant driver of security-optimal detection performance.}
\label{fig:reward_comparison}
\end{center}

To isolate the causal effect of reward design with all other variables fixed, DDQN was evaluated under the same Seed 456 and $\gamma=0.1$ setting with only the reward function varied. As shown in Table~\ref{tab:reward_ablation}, $R_2$ increased recall from 0.9860 to 0.9920 and reduced the FNR from 0.0140 to 0.0080, corresponding to a 43\% relative reduction. $R_1$ yielded slightly higher precision (0.9930 vs. 0.9910) and lower FPR (0.0070 vs. 0.0090); however, these differences were operationally negligible: a false positive generated a reversible quarantine alert, whereas a false negative allowed ransomware execution and irreversible data encryption. This controlled ablation confirms that the asymmetric reward structure itself, rather than $\gamma$ selection or seed variation, is the primary causal driver of the observed FNR improvement.

\begin{table*}[!t]
\caption{Effect of reward function on DDQN performance under Seed 456 and $\gamma=0.1$.}
\label{tab:reward_ablation}
\centering
\footnotesize
\setlength{\tabcolsep}{5pt}
\resizebox{\textwidth}{!}{%
\begin{tabular}{lccccccp{5.2cm}}
\toprule
\textbf{Reward} &
\textbf{Accuracy} &
\textbf{Precision} &
\textbf{Recall} &
\textbf{F1 Score} &
\textbf{FNR} &
\textbf{FPR} &
\textbf{Observation} \\
\midrule
$R_1$ &
$0.9895 \pm 0.0037$ &
$0.9930 \pm 0.0027$ &
$0.9860 \pm 0.0065$ &
$0.9895 \pm 0.0037$ &
$0.0140 \pm 0.0065$ &
$0.0070 \pm 0.0027$ &
Lower false alarms, but higher missed-attack risk \\
$R_2$ &
$0.9915 \pm 0.0038$ &
$0.9910 \pm 0.0041$ &
$0.9920 \pm 0.0057$ &
$0.9915 \pm 0.0038$ &
$0.0080 \pm 0.0057$ &
$0.0090 \pm 0.0042$ &
Best security-oriented performance \\
\bottomrule
\end{tabular}%
}
\end{table*}

To statistically confirm that the security gains attributed to $R_2$ over $R_1$ are not artifacts of stochastic training variance, per-algorithm Wilcoxon signed-rank tests were conducted on the mean FNR distributions across all seeds and $\gamma$ values ($n=12$ paired observations per algorithm: four $\gamma$ values $\times$ three seeds). Table~\ref{tab:wilcoxon_reward} reports the test statistics. Across all four agents, $R_2$ achieved a statistically significantly lower FNR than $R_1$, confirming the causal effect of the asymmetric reward design on missed-detection minimization. The strongest effects were observed for DDQN and PPO ($p<0.001$), whereas A2C achieved significance at $p<0.05$, consistent with its comparatively unstable optimization behavior. Pooled across all agents, the omnibus test confirmed $R_2$'s superiority at $p<0.0001$. These results establish that the $R_2$ reward function, rather than hyperparameter fluctuation or random seed variation, is the dominant factor driving the security-aligned detection behavior observed throughout this study.

\begin{table*}[!t]
\caption{Wilcoxon signed-rank test results for $R_1$ vs. $R_2$ mean FNR comparison per algorithm ($n=12$ paired observations per algorithm across four $\gamma$ values and three seeds). $R_2$ statistically significantly outperformed $R_1$ across all four agents and in the pooled analysis.}
\label{tab:wilcoxon_reward}
\centering
\footnotesize
\setlength{\tabcolsep}{5pt}
\resizebox{\textwidth}{!}{%
\begin{tabular}{lccccccc}
\toprule
\textbf{Algorithm} &
\textbf{$R_1$ Mean FNR} &
\textbf{$R_2$ Mean FNR} &
\textbf{$\Delta$FNR ($R_1-R_2$)} &
\textbf{W-statistic} &
\textbf{$p$-value} &
\textbf{Significance} &
\textbf{Verdict} \\
\midrule
DQN & 0.0138 & 0.0106 & +0.0031 & 2.5 & 0.0156 & * & $R_2$ wins \\
DDQN & 0.0200 & 0.0135 & +0.0065 & 0.0 & 0.0005 & *** & $R_2$ wins \\
PPO & 0.0392 & 0.0250 & +0.0142 & 0.0 & 0.0005 & *** & $R_2$ wins \\
A2C & 0.0473 & 0.0396 & +0.0077 & 7.0 & 0.0332 & * & $R_2$ wins \\
ALL (pooled) & 0.0301 & 0.0222 & +0.0079 & 40.5 & $<0.0001$ & *** & $R_2$ wins \\
\bottomrule
\end{tabular}%
}

\vspace{1mm}
\footnotesize
\textit{Significance codes:} *** $p<0.001$, * $p<0.05$. $\Delta$FNR values represent the absolute reduction in mean FNR achieved by $R_2$ relative to $R_1$. The W-statistic is the Wilcoxon signed-rank W value; smaller W indicates stronger directional dominance of $R_2$.
\end{table*}

\subsection{Effect of Discount Factor $\gamma$ (RQ2)}
\label{subsec:discount_factor_effect}

The FNR heatmap shown in Fig.~\ref{fig:fnr_heatmap} reveals that $\gamma$ operates as a security-critical hyperparameter with strong agent-specific effects. Across the seed-specific best configurations (Tables~\ref{tab:seed42_results}--\ref{tab:seed456_results}), lower discount factors, particularly $\gamma=0.10$ and $\gamma=0.50$, are most frequently associated with the strongest security performance, whereas higher values ($\gamma=0.90$ and $\gamma=0.99$) appear less often among top-ranked configurations.

\begin{figure*}[!t]
\centering
\includegraphics[width=0.98\textwidth]{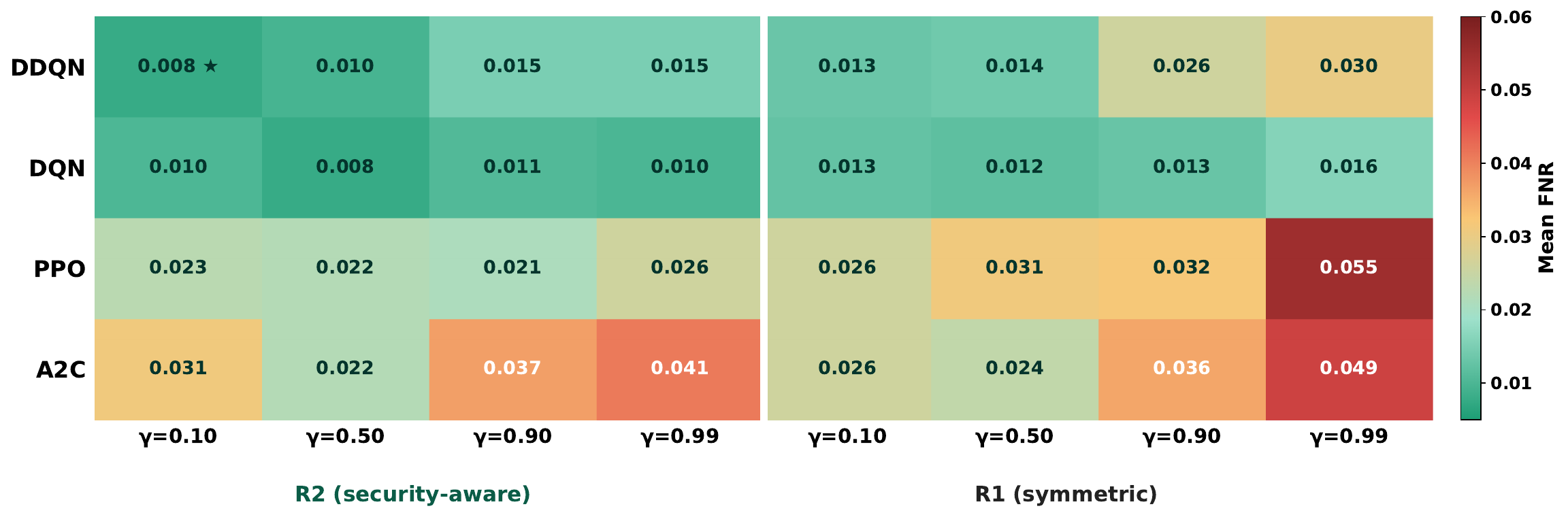}
\caption{Mean FNR across all 32 configurations (four agents $\times$ two rewards $\times$ four discount factors), averaged over five-fold cross-validation and three seeds. Green cells indicate low missed-detection risk, whereas red cells indicate high risk. $R_2$ configurations consistently produce lower FNR than $R_1$ configurations, and high $\gamma$ values systematically increase FNR for policy-gradient agents.}
\label{fig:fnr_heatmap}
\end{figure*}

Smaller discount factors prioritize immediate classification outcomes over long-horizon cumulative reward estimation. Because ransomware classification decisions are largely independent and operational penalties are immediate, lower $\gamma$ values ($0.10$--$0.50$) appear more suitable than larger $\gamma$ settings ($0.90$--$0.99$), which may amplify delayed reward-estimation noise without improving security-oriented decision quality.

The value-based agents (DQN and DDQN) remained largely stable across all four $\gamma$ values. As shown in Fig.~\ref{fig:fnr_heatmap}, DDQN's FNR ranges from 0.008 ($\gamma=0.10$) to 0.015 ($\gamma=0.99$) under $R_2$, a 72\% relative increase, yet remains operationally acceptable throughout. DQN showed even lower $\gamma$ sensitivity, with FNR values ranging from 0.008 to 0.011. This stability arises because off-policy experience replay partly decouples the effective planning horizon from the discount structure, while the episodic i.i.d. nature of the detection task means that long-horizon dependencies provide limited additional benefit.

Policy-gradient agents (PPO and A2C) were substantially more sensitive. Under $R_2$, A2C degraded from 0.031 ($\gamma=0.10$) to 0.041 ($\gamma=0.99$), corresponding to a 32\% relative increase. The degradation was more severe under $R_1$: A2C's FNR increased from 0.026 ($\gamma=0.10$) to 0.049 ($\gamma=0.99$), while PPO reached 0.055 at $\gamma=0.99$, representing the darkest red cell in Fig.~\ref{fig:fnr_heatmap} and an operationally unacceptable missed-detection rate for ransomware defense. This behavior likely reflects unstable truncated-return estimation under A2C's short rollout horizon ($n_{\mathrm{steps}}=5$), where high $\gamma$ values amplify accumulated value-estimation noise and destabilize policy optimization. The heatmap makes this trend visually explicit: $R_2$ cells are consistently greener than their $R_1$ counterparts at every $\gamma$ level, and cell color darkens systematically from left ($\gamma=0.10$) to right ($\gamma=0.99$) for policy-gradient agents.

In summary, $\gamma=0.10$ minimizes missed-detection risk across all agents. High $\gamma$ increases FNR by up to 89\% for A2C under $R_1$, directly establishing the discount factor as a security-critical deployment hyperparameter rather than a routine RL tuning parameter.

\subsection{Statistical Analysis}
\label{subsec:statistical_analysis}

Fig.~\ref{fig:statistical_validation} presents the statistical validation results. Panel (a) shows the Wilcoxon signed-rank post-hoc test results for all six pairwise agent comparisons, plotted as $-\log_{10}(\textit{p-value})$ horizontal bars. The red dashed vertical line marks the $p=0.001$ threshold. All six bars extend well beyond this threshold, confirming that every pairwise performance difference is statistically significant at $p<0.001$. The two shortest bars --- DDQN vs.\ DQN ($p=0.00065$) and A2C vs.\ PPO ($p=0.000494$) --- represent the closest-performing pairs, yet both remain statistically significant at $p<0.001$.

Panel (b) presents the complete statistical summary, confirming that the Friedman test globally rejects the null hypothesis of equivalent performance ($\chi^2 = 61.30$, $p<0.001$). The Wilcoxon post-hoc tests further establish the strict performance ordering DDQN $\succ$ DQN $\succ$ PPO $\succ$ A2C across all paired comparisons. Together, these results provide strong non-parametric statistical evidence that the observed performance differences are not attributable to stochastic training variance.

\begin{figure*}[!t]
\centering
\includegraphics[width=0.98\textwidth]{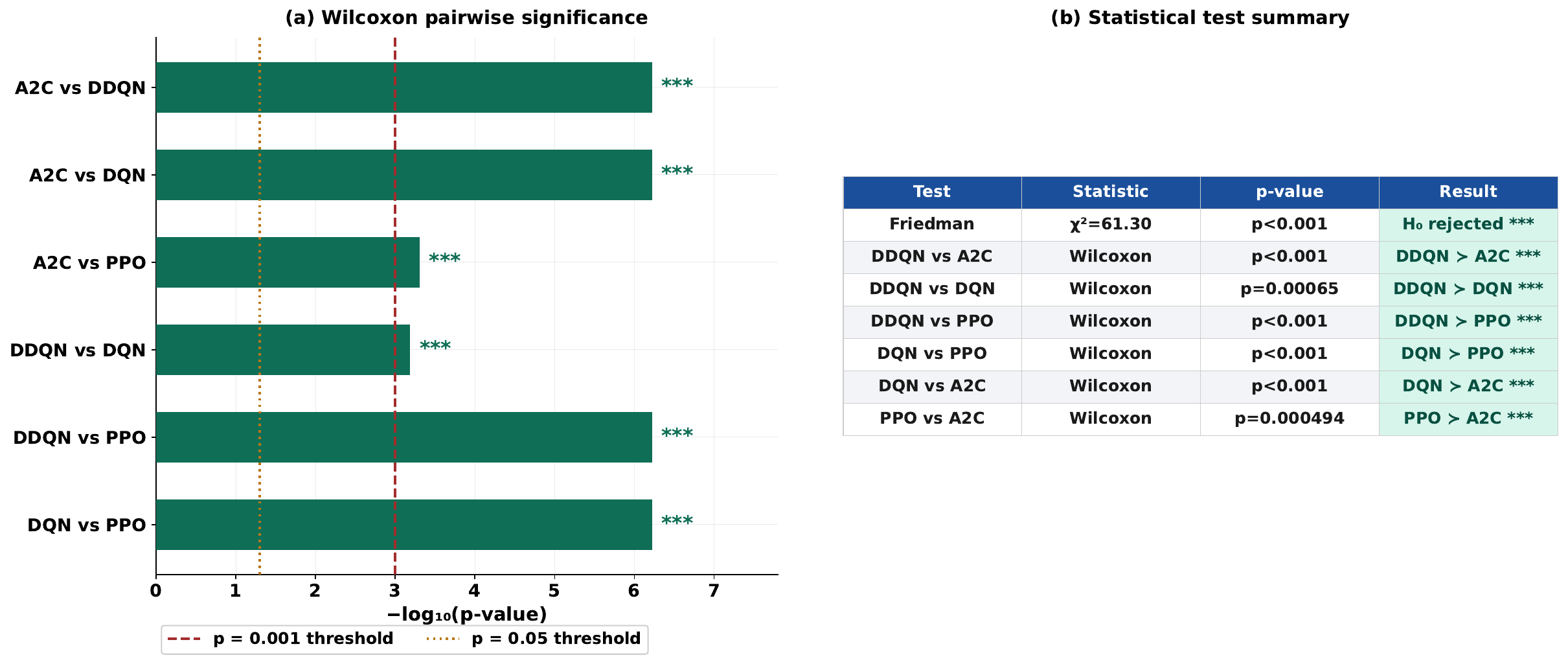}
\caption{Statistical validation of DRL agent performance. Panel (a) shows Wilcoxon signed-rank post-hoc comparisons plotted as $-\log_{10}(p)$ values for all six pairwise comparisons. The dashed vertical line indicates the $p=0.001$ significance threshold. Panel (b) summarizes the Friedman omnibus test and Wilcoxon post-hoc results, confirming statistically significant performance differences across all DRL agents.}
\label{fig:statistical_validation}
\end{figure*}

\subsection{Comparison with State-of-the-Art Related Work}
\label{subsec:sota_comparison}

To contextualize the effectiveness of the proposed SA-DRL framework, Table~\ref{tab:sota_comparison} compares the SOMS-selected configuration (DDQN under $R_2$ with $\gamma=0.1$) with representative prior work spanning static analysis, behavioral ML/DL, cost-sensitive supervised learning, and DRL-based detection. The comparison highlights three principal distinctions of the proposed framework: (i) it is the first ransomware detection framework to embed asymmetric FN--FP operational costs directly into a DRL reward signal operating on behavioral runtime telemetry; (ii) it achieves the lowest reported FNR (0.80\%) among all compared approaches; and (iii) it provides the only security-aware model selection criterion (SOMS) validated by non-parametric statistical tests across 480 controlled training runs.

\begin{table*}[!t]
\caption{Comparison of SA-DRL with state-of-the-art ransomware detection methods and DRL-based security systems that inspired the asymmetric reward design. ``--'' = metric not reported. $\dagger$ = FNR derived from reported recall ($\textnormal{FNR}=1-\textnormal{Recall}$). * = non-ransomware domain (network intrusion detection); included to show that asymmetric DRL reward design existed in adjacent security domains but had not been applied to behavioral ransomware detection prior to this work. The best result per column is shown in bold.}
\label{tab:sota_comparison}

\centering
\scriptsize
\setlength{\tabcolsep}{3pt}

\resizebox{\textwidth}{!}{%
\begin{tabular}{p{1.2cm} p{2.9cm} p{2.6cm} p{2.3cm} c c c p{2.5cm} p{2.6cm}}
\toprule

\textbf{Study} &
\textbf{Method} &
\textbf{Feature Type} &
\textbf{Cost-Aware Training} &
\textbf{Accuracy} &
\textbf{F1} &
\textbf{FNR (\%)} &
\textbf{Statistical Validation} &
\textbf{Security Selection} \\

\midrule

\multicolumn{9}{c}{\textbf{Ransomware Detection Methods}} \\

\midrule

~\cite{ref31} &
RF, GB Trees &
Behavioral (dynamic) &
No &
$\sim$99\% &
$\sim$0.990 &
-- &
No &
No \\

~\cite{ref32} &
RF (DNA seq.) &
Behavioral API calls &
No &
97.3\% &
-- &
-- &
No &
No \\

~\cite{ref15} &
API call sequences &
No &
98.5\% &
0.985 &
1.5$\dagger$ &
No &
No \\

~\cite{ref9} &
Cost-sensitive Pareto Ensemble &
Behavioral + static &
Post-hoc only &
98.5\% &
0.985 &
-- &
No &
No \\

~\cite{ref33} &
XAI deep learning &
Behavioral (dynamic) &
No &
98.1\% &
0.980 &
-- &
No &
No \\

~\cite{ref18} &
BERT/RoBERTa &
API call sequences &
No &
99.0\% &
0.990 &
1.0$\dagger$ &
No &
No \\

~\cite{ref16} &
DRL (symmetric $R$) &
Static (PE header) &
No &
97.6\% &
0.976 &
-- &
No &
No \\

\midrule

\multicolumn{9}{c}{\textbf{DRL Security --- Adjacent Domains (Asymmetric Reward Precedent)*}} \\

\midrule

~\cite{ref19}* &
Dueling DDQN (10:1 asym.) &
Network traffic (IDS) &
Embedded in reward &
$\sim$98\% &
-- &
-- &
No &
No \\

~\cite{ref38}* &
Reward shaping (DRL) &
Network security &
Embedded in reward &
-- &
-- &
-- &
No &
No \\

\midrule

\multicolumn{9}{c}{\textbf{Proposed Framework}} \\

\midrule

\textbf{SA-DRL} &
\textbf{DDQN + Asym.\ $R_2$ (4:1)} &
\textbf{Behavioral (runtime)} &
\textbf{Embedded in reward} &
\textbf{99.15\%} &
\textbf{0.9915} &
\textbf{0.80} &
\textbf{Friedman + Wilcoxon} &
\textbf{SOMS (security-first)} \\

\bottomrule
\end{tabular}

}
\end{table*}

Table~\ref{tab:sota_comparison} reveals three layers of structural distinction. Among ransomware detection methods, behavioral ML/DL approaches~\cite{ref8,ref15,ref31} achieve 97--99\% accuracy but optimize symmetric loss functions, leaving the FN--FP cost asymmetry unaddressed; their reported or derived FNR values (1.0--1.5\%) remain consistently higher than SA-DRL's 0.80\%. Deng et al.~\cite{ref16}, the only prior work applying DRL to ransomware classification, evaluate the same four agents (DQN, DDQN, PPO, and A2C) but limit discount-factor analysis to two values ($\lambda=0.01,0.99$), rely on static PE-header features vulnerable to binary obfuscation, and apply a symmetric reward formulation (correct = 1, incorrect = 0) that leaves FN--FP cost asymmetry entirely unaddressed. SA-DRL extends this foundation through four discount factors, asymmetric security-aware reward functions, five-fold cross-validation, and three-seed replication, yielding 480 controlled runs validated by Friedman and Wilcoxon tests under a formal security-first selection criterion (SOMS) absent from~\cite{ref16}.

Zahoora et al.'s cost-sensitive Pareto Ensemble~\cite{ref9} is the only prior ransomware work to directly target cost asymmetry, yet applies it post-hoc to a pre-trained classifier rather than embedding it directly into the learning objective, a structural distinction that directly motivates the DRL reward-shaping approach. Among DRL security methods, Haddane et al.~\cite{ref19} and Bates \& Hicks~\cite{ref38} established that asymmetric reward design produces security-aligned behavior in intrusion detection, but neither applied this principle to behavioral ransomware telemetry. SA-DRL therefore represents the first framework to unify asymmetric reward shaping, behavioral runtime ransomware analysis, discount-factor security analysis, multi-agent DRL comparison, and statistically validated security-first model selection within a single ransomware detection framework.

\section{Discussion}
\label{sec:discussion}

This study demonstrates that embedding asymmetric security costs directly into the reward signal enables DRL agents to learn ransomware detection policies that align more effectively with operational cybersecurity priorities. The principal finding is that the security-aware reward function $R_2$ consistently reduced the false-negative risk relative to the symmetric reward $R_1$. Under the SOMS-selected configuration, DDQN with $R_2$ and $\gamma = 0.1$ achieved the lowest missed-detection rate while maintaining strong F1-score performance and computational efficiency.

The results confirm that reward design is not merely a tuning choice in DRL-based ransomware detection but a central mechanism for aligning learning behaviour with security objectives. Prior supervised ransomware detectors based on static or behavioral features and deep learning achieved high post-hoc accuracy~\cite{ref5,ref27,ref8,ref9,ref13,ref14,ref15}, but most optimize symmetric loss functions in which false positives and false negatives contribute equally to the objective. This is unsuitable for ransomware defense because a false negative may permit encryption, data loss, service disruption, and financial damage, whereas a false positive usually results in a reversible alert or quarantine. This aligns with recent evidence showing that ransomware recovery costs and operational disruption remain severe~\cite{ref1,ref2,ref3,ref4}, supporting the need to move beyond accuracy-centered evaluation toward security-aware optimization. By assigning a larger penalty to missed ransomware, the proposed $R_2$ shifts the agent’s optimization pressure toward reducing the most operationally harmful error types. This differs fundamentally from post-hoc cost-sensitive approaches such as Zahoora et al.~\cite{ref9}, where cost adjustment is applied after classifier construction. In SA-DRL, the agent is exposed to the security cost during training, enabling policy learning to be shaped by missed-detection risk from the first update.

Some consistent patterns emerged across the 480 controlled training runs. First, the asymmetric reward design functions as a primary security-control mechanism rather than a secondary optimization choice. By explicitly encoding FN--FP asymmetry into the learning objective (Table~\ref{tab:reward_ablation}), the proposed $R_2$ reduced the mean FNR by 43\% relative to $R_1$, whereas the FPR increased by only 0.20 percentage points, an operationally negligible trade-off because false positives generate reversible alerts, whereas false negatives permit irreversible ransomware encryption. The Wilcoxon signed-rank tests (Table~\ref{tab:wilcoxon_reward}) confirm that this effect is statistically significant for all four agents (all $p \leq 0.033$) and remains robust across seed and discount-factor variations (pooled $p < 0.0001$). These findings support Ibrahim et al.'s~\cite{ref20} theory that asymmetric reward structures improve safety-critical classification performance.

The comparison between value-based and policy-gradient agents further indicates that DRL architectures respond differently to asymmetric reward shaping. DQN and DDQN remained comparatively stable across discount factors, whereas PPO and A2C exhibited substantially greater sensitivity, particularly at higher $\gamma$ values. The statistically validated hierarchy (Fig.~\ref{fig:statistical_validation}), DDQN $\succ$ DQN $\succ$ PPO $\succ$ A2C (Friedman $\chi^2 = 61.30$, $p < 0.001$; all Wilcoxon pairwise tests $p < 0.001$), aligns with prior studies showing that value-based methods excel in discrete classification environments~\cite{ref19,ref35}. DDQN's advantage over DQN arises from its more stable Q-value estimation, which reduces overestimation errors that can increase missed ransomware detections~\cite{ref37}. This distinction is particularly important in security-sensitive environments, as Merzouk et al.~\cite{ref34} similarly reported that A2C exhibits the highest false-negative rates under adversarial perturbation, a trend reproduced in this study.

The effect of the discount factor further suggests that ransomware detection in this setting is governed primarily by immediate classification consequences rather than long-horizon environmental control. Lower $\gamma$ values consistently produced stronger security-oriented performance because each state represented an independent behavioral sample and the security consequence was determined immediately after classification. Smaller discount factors therefore prioritize immediate detection quality over delayed cumulative reward estimation. This finding aligns with prior DRL security studies showing that discount-factor selection can substantially influence detection behavior~\cite{ref35}, particularly in classification-oriented environments.

The proposed SOMS criterion provides a more deployment-relevant model-selection rule than aggregate accuracy alone. In ransomware defense, two models with similar accuracy may produce substantially different operational risks if one exhibits a higher false-negative rate. By prioritizing minimum FNR, followed by maximum F1-score and minimum training time, SOMS aligns model selection with the practical objective of preventing undetected ransomware execution. This is particularly important because many prior studies report high overall accuracy~\cite{ref7,ref8,ref13,ref14,ref15} while giving comparatively limited attention to the asymmetric operational consequences of different error types.

The episode-level permutation mechanism also contributes to the framework by reducing order-induced bias and enabling implicit adaptive sample emphasis. Difficult ransomware samples that repeatedly trigger high FN penalties receive stronger corrective updates across varying episode positions. This provides a lightweight alternative to explicit reweighting approaches such as class-weighted loss while remaining consistent with reward-shaping theory, where the magnitude and frequency of reward signals influence learned behavior~\cite{ref17,ref20}.

From a deployment perspective, SA-DRL is most suitable for sandbox-assisted ransomware triage, behavioral endpoint monitoring, and high-risk environments in which missed ransomware is substantially more costly than additional false alarms. The conservative 4:1 FN--FP penalty used in $R_2$ is intentionally less aggressive than some prior DRL security frameworks~\cite{ref18}, yet still sufficient to produce statistically significant reductions in missed detections. The selected ratio provides a stable trade-off for the balanced dataset used in this study while remaining flexible for recalibration under different class priors, alert-fatigue constraints, or sector-specific operational risks. For example, healthcare or critical-infrastructure environments may justify stronger FN penalties, whereas high-volume enterprise SOC deployments may require tighter FPR control.

Several limitations remain. First, the evaluation is confined to a single balanced dataset of 2,000 samples; the relative effectiveness of $R_1$ and $R_2$, as well as the suitability of the 4:1 penalty ratio, may differ under severe class imbalance or substantially larger corpora. Second, the study focuses on binary ransomware detection rather than ransomware family attribution or pre-encryption early warning. Third, although the experiments include multiple agents, rewards, discount factors, folds, and seeds, the evaluation remains offline and does not model continual adaptation against evolving ransomware behavior. Finally, the reward ratio used in $R_2$ was intentionally conservative and empirically effective in this setting, but it should not be interpreted as universally optimal across all operational environments. Future work should therefore investigate online reinforcement learning over live telemetry streams, multi-source class-imbalanced datasets, adaptive reward recalibration under concept drift, and extension toward early-stage ransomware family prediction.

Overall, the findings demonstrate that security-aware reward design can shift DRL-based ransomware detection from general accuracy optimization toward explicit missed-detection minimization. DDQN with asymmetric reward shaping and low discounting emerged as the most suitable configuration under SOMS, suggesting that future ransomware detectors should be evaluated not only by accuracy or F1-score, but also by whether their learning objectives and model-selection strategies reflect the asymmetric operational realities of cybersecurity defense.

\section{Conclusion and Future Work}
\label{sec:conclusion}

This study proposed a security-aware deep reinforcement learning (SA-DRL) framework for behavioral ransomware detection by incorporating asymmetric security costs into the reward function. It addressed a key limitation of conventional machine learning approaches that treat false negatives and false positives equally despite the substantially higher operational impact of missed ransomware attacks. The framework also addressed behaviorally diverse ransomware samples through adaptive policy learning. Across 480 controlled experiments, value-based DRL methods consistently outperformed policy-gradient approaches, with DDQN under the asymmetric reward setting ($R_2$) and $\gamma = 0.1$ achieving the strongest overall performance. The optimal configuration achieved an F1-score of 0.9915 and a false-negative rate of 0.0080, reducing missed ransomware detections by 67.6\% relative to the best supervised baseline. The results demonstrate that asymmetric reward shaping produces more stable and security-aligned detection behavior than symmetric optimization objectives. Overall, the findings confirm that cost-aware reinforcement learning can effectively align ransomware detection with operational cybersecurity priorities while handling behaviorally heterogeneous samples. Although evaluated on a binary classification dataset, the approach shows strong potential for real-world malware detection. During deployment, only forward-pass inference is required, introducing no additional online reinforcement learning overhead.

Several directions remain for future research. First, the evaluation should be extended to larger and naturally imbalanced telemetry streams to assess robustness under operational deployment conditions. Second, future studies should investigate continual and online DRL adaptation for evolving ransomware behaviors and concept drift. Third, ransomware family-level classification, pre-encryption early warning, and adversarial robustness against evasion-aware ransomware remain important extensions. Finally, adaptive or dynamically calibrated reward ratios may further improve security alignment across different sectors, threat models, and alert-tolerance requirements.








\printcredits

\section*{Supplementary Material}

Complete per-configuration experimental results across all
algorithms, reward functions, discount factors, and random seeds
are provided in the Supplementary Material submitted alongside
this manuscript.

\bibliographystyle{cas-model2-names}

\bibliography{references}



\end{document}